\documentclass[aps,prl,preprint,superscriptaddress]{revtex4-1}

\usepackage{hyperref}
\usepackage{graphicx} 
\DeclareGraphicsExtensions{.pdf,.png,.jpg}

\usepackage{amsmath}
\setcitestyle{super}

\usepackage[normalem]{ulem} 
\usepackage{color}

\newcommand{\blue}[1]{{\color{blue} {#1}}}

\begin{document}


\title{The importance of the weak: Interaction modifiers in artificial spin ices}




\author{Erik \"{O}stman}
\email{erik.ostman@physics.uu.se}
\affiliation{%
 Department of Physics and Astronomy, Uppsala University, Box 516, SE-75120, Uppsala, Sweden
}%
\author{Henry Stopfel}%
\affiliation{%
 Department of Physics and Astronomy, Uppsala University, Box 516, SE-75120, Uppsala, Sweden
}%
\author{Ioan-Augustin Chioar}%
\affiliation{%
 Department of Physics and Astronomy, Uppsala University, Box 516, SE-75120, Uppsala, Sweden
}%
\author{Unnar B. Arnalds}%
\affiliation{%
 Science Institute, University of Iceland, Dunhaga 3, 107 Reykjavik, Iceland
}%
\author{Aaron Stein}%
\affiliation{%
 Center for Functional Nanomaterials, Brookhaven National Laboratory, Upton, New York 11973, USA
}%
\author{Vassilios Kapaklis}%
\affiliation{%
 Department of Physics and Astronomy, Uppsala University, Box 516, SE-75120, Uppsala, Sweden
}%
\author{Bj\"{o}rgvin Hj\"{o}rvarsson}%
\affiliation{%
 Department of Physics and Astronomy, Uppsala University, Box 516, SE-75120, Uppsala, Sweden
}%

\date{\today}
\maketitle

{\bf
The modification of geometry and interactions in two-dimensional magnetic nanosystems has enabled a range of studies addressing the magnetic order \cite{Wang2006,Mengotti2010,Morgan2010,Kapaklis_2012_NJP,Gilbert_2014_NatPhys,Gilbert_2016_NatPhys,Gilbert_2014_NatPhys}, collective low-energy dynamics \cite{Farhan_2013_NatPhys,Vassilios2014}, and emergent magnetic properties \cite{Gilbert_2014_NatPhys,Perrin2016,Canals_2016dc}, in e.g. artificial spin ice structures. The common denominator of all these investigations is the use of Ising-like mesospins as building blocks, in the form of elongated magnetic islands.  Here we introduce a new approach: single interaction modifiers, using slave-mesospins in the form of discs, within which the mesospin is free to rotate in the disc plane\cite{Unnar2014}. We show that by placing these on the vertices of square artificial spin ice arrays and varying their diameter, it is possible to tailor the strength and the ratio of the interaction energies. We demonstrate the existence of degenerate ice-rule obeying states in square artificial spin ice structures, enabling the exploration of thermal dynamics in a spin liquid manifold. Furthermore, we even observe the emergence of flux lattices on larger length-scales, when the energy landscape of the vertices is reversed. The work highlights the potential of a design strategy for two-dimensional magnetic nano-architectures, through which mixed dimensionality of mesospins can be used to promote thermally emergent mesoscale magnetic states.
}

\begin{figure}[h]
\includegraphics[width=0.8\columnwidth]{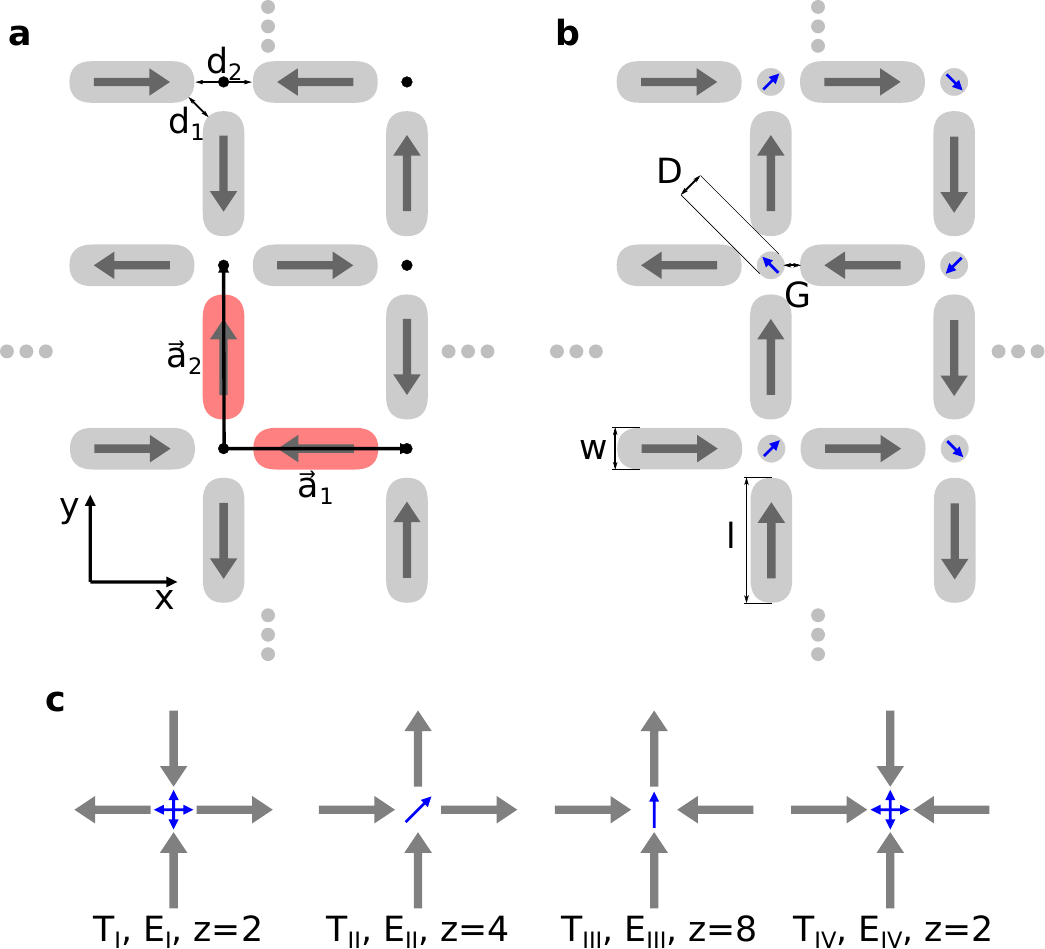}
\caption{\label{fig:intro} \textbf{Schematics of the SASI and mSASI lattices shown together with their vertex types}. \textbf{a}, Illustration of a SASI and \textbf{b} a mSASI lattice. Real space base vectors defining the square lattice together with the base (red islands), are shown in \textbf{a} together with the discrete lattice points (black dots). The lattice parameter $\alpha$ is defined as $\alpha=||\mathbf{a}_1||=||\mathbf{a}_2||$. The length of the islands is l=450 nm and their width is w=150 nm. \textbf{c}, The four different vertex types with their respective energies, E, and degeneracy, z. In T$_{\text{II}}$ and T$_{\text{III}}$ vertices the XY-like spins of the discs have a well defined direction due to the residual stray field. In T$_{\text{I}}$ and T$_{\text{IV}}$ vertices the XY spins exhibit weak fourfold degeneracy (in the collinear approximation\cite{Unnar2014}).}
\end{figure}

Lithographic techniques can be used to fabricate magnetic nano-arrays, in which the interaction between the elements can be chosen by e.g. the distance between the islands. This approach has been used in a number of previous works, addressing both the order and dynamics of magnetic nanostructures \cite{Wang2006,Mengotti2010,Morgan2010,Kapaklis_2012_NJP,Gilbert_2014_NatPhys,Gilbert_2016_NatPhys,Gilbert_2014_NatPhys, Farhan_2013_NatPhys,Vassilios2014, Arnalds2012}. In the specific case of square artificial spin ice (SASI) this approach has even enabled tailoring of the thermal dynamics and relaxation\cite{Vassilios2014, Andersson_2016_SciRep,Morley_PRB_2017,Farhan2013},
as well as experimental realizations\cite{Perrin2016} of the degenerate square-ice model\cite{Lieb1967}.
The distance and thereby the coupling strength for nearest and next-nearest neighbours are different in SASI (d$_1\neq\text{d}_2$ (see Fig. \ref{fig:intro})), resulting in the loss of degeneracy.
As a consequence, the ice-rule obeying vertices, with two islands pointing in - two islands pointing out, are split into two groups (T$_{\text{I}}$ and T$_{\text{II}}$) with different energies (E$_{\text{I}}<\text{E}_{\text{II}}$). 
One way to remedy this shortcoming is to shift parts of the lattice in the third dimension\cite{Moller2006,Chern2014, Perrin2016}. 
An alternative way to modify the energy landscape, is to introduce an interaction modifier, as illustrated in Fig. \ref{fig:intro}b.  In these modified SASI (mSASI) arrays, all islands have the same distance, or gap G, to the interaction modifier. While a height offset might seem as the obvious choice for manipulating the coupling strengths between the islands, the use of interaction modifiers at the vertices of artificial spin ice structures is not only lithographically much easier to obtain, but also opens up completely new avenues for tailoring their energy landscapes. Instead of having a system consisting of only one type of islands, we use two sub-systems with widely different  shape anisotropies and activation energies. 

The elongated islands used in artificial spin ice structures can be described as Ising-like mesospins, while the discs we use here to modify their interaction, can be described to a first approximation as XY-like\cite{Unnar2014}. The difference in their activation energies will  give rise to a master-slave relation, where the vertex state dictates the direction of the XY-mesospin. The mutual interaction of the Ising and the XY spins yields the emergent magnetic order. 
For T$_{\text{II}}$ and T$_{\text{III}}$ vertices the magnetization direction of the enclosed discs is enforced by the effective dipole moment of the vertex but the magnetic state of the disc for T$_{\text{I}}$ and T$_{\text{IV}}$ vertices are fourfold degenerate (see Supplementary Fig. 14).  
By fabricating the islands from a material with an ordering temperature at or below room temperature \cite{Arnalds2012, Vassilios2014, Andersson_2016_SciRep} allows us to access all the relevant parts of the phase diagram: from the paramagnetic state of the material to the ordering of the vertices as described below. 
 
\begin{figure}[h!]
\includegraphics[width=0.45\columnwidth]{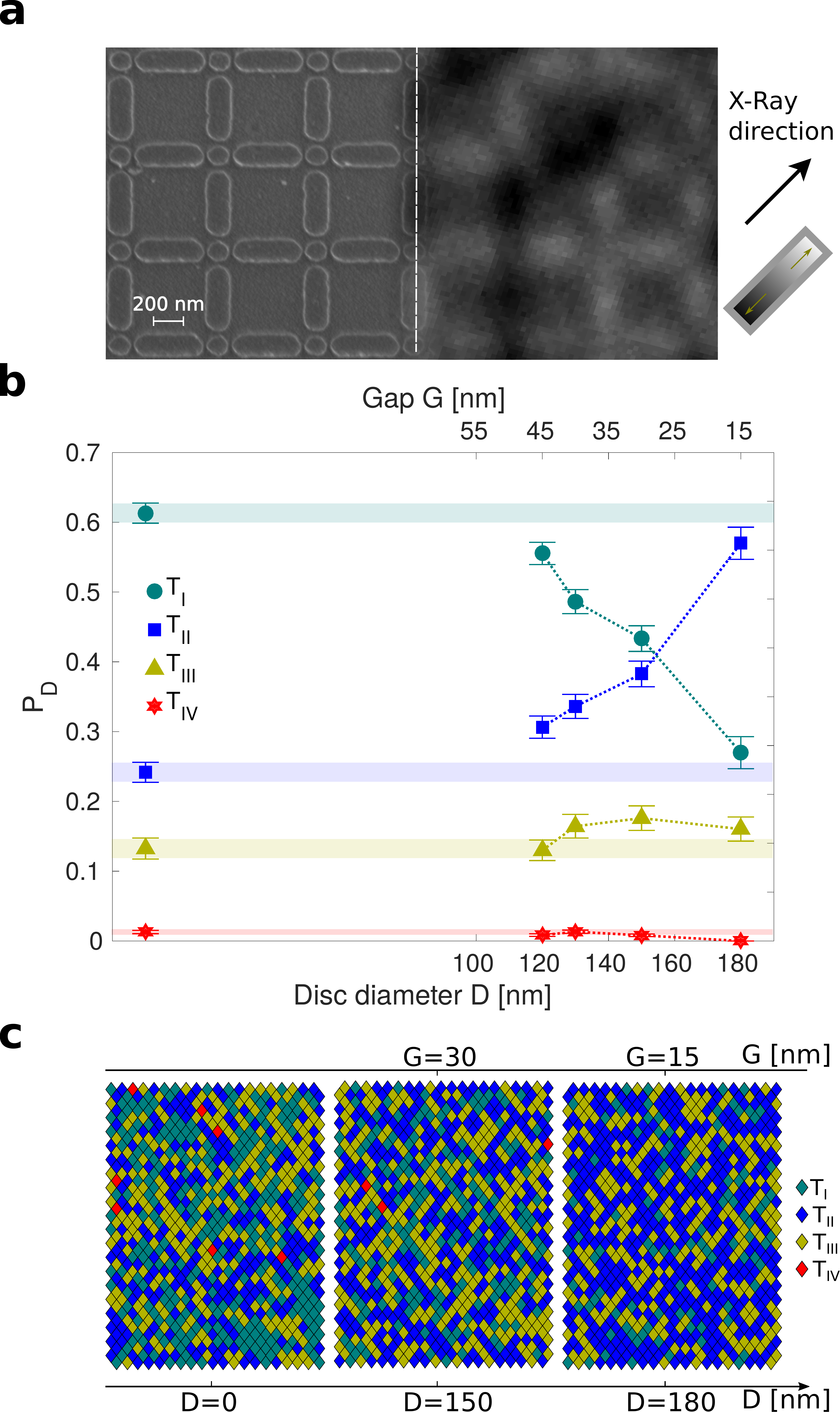}
\caption{\label{fig:vpopulation}\textbf{Real space magnetic imaging and vertex populations}. \textbf{a}, Representative SEM and PEEM-XMCD images ($\alpha$=660 nm, D=130 nm and G=40). X-ray direction and color contrast is indicated by the arrows. \textbf{b}, Degeneracy normalized vertex populations P$_\text{D}$ as a function of disc diameter D. The horizontal coloured stripes refer to the populations for D=0 and serve as a guide to the eye along with the dashed lines, highlighting the vertex population changes while varying D. As the diameter D increases, the population ratios between T$_{\text{I}}$ and T$_{\text{II}}$ are reversed, pointing towards a transition in vertex energies from E$_{\text{I}}<\text{E}_{\text{II}}$ to E$_{\text{I}}>\text{E}_{\text{II}}$, where D=150 nm is closest realization to E$_{\text{I}}$=E$_{\text{II}}$ condition. E$_{\text{IV}}> \text{E}_{ \text{III}}> (\text{E}_{\text{I}}$, E$_{\text{II}}$) holds for all lattices explored. \textbf{c}, Spatial representation, in the form of vertex maps, of the data from \textbf{b} for D=0, D=150, and D=180. The color coding represents the four different types of vertices, described in Fig. \ref{fig:intro}c, using the colorscheme from Fig. \ref{fig:vpopulation}b.}
\end{figure}

We have studied in total 15 different SASI arrays with three different lattice parameters $\alpha$=[660,720,800] nm patterned on $\delta$-doped Pd(Fe) thin films. Each array has five different disc sizes, D=[0,120,130,150,180] nm for $\alpha$=660 nm and D=[0,130,150,180,200] nm for $\alpha$=[720,800] nm. All arrays have the same elongated island size of 450 $\times$ 150 nm$^2$.  Photoemission Electron Microscopy (PEEM) based upon X-ray Magnetic Circular Dichroism (XMCD) was utilised for determining the magnetic state of the elements. Representative results from the near-perturbation free measurements are shown in Fig. \ref{fig:vpopulation}a.

The degeneracy normalized vertex populations, derived from the PEEM-XMCD images, as a function of disc diameter for the $\alpha$=660 nm array are presented in Fig. \ref{fig:vpopulation}b (see Supplementary Fig. 1 for full data set). The sample is thermally active and upon cooling will pass the blocking temperatures of the islands and the discs. This implies that the frozen states observed at 120 K represent configurations which were thermally arrested at higher temperatures\cite{Vassilios2014}. In the absence of a central disc, a large number of excitations are observed as the system approaches the antiferromagnetically-ordered ground state of SASI\cite{Morgan2010}. The high degree of disorder is caused by the relatively small coupling strength between the Ising islands. 

The presence of the discs dramatically changes the overall behavior as seen in Fig. \ref{fig:vpopulation}b, passing from a T$_{\text{I}}$ dominated spin texture in the absence of discs (D=0) to a T$_{\text{II}}$ dominated configuration when discs with a diameter of 180 nm are present. This implies an inversion in the energy levels for E$_{\text{I}}$ and E$_{\text{II}}$, at a diameter around 150 nm, corresponding to a gap of 30 nm (see Supplementary Fig. 13 for micro-magnetic simulations). We also note that the populations of T$_{\text{III}}$ and T$_{\text{IV}}$ appear to be only weakly affected by the presence of the discs, within the range of diameters studied here (see Fig. \ref{fig:vpopulation}). Therefore the corresponding vertex energies remain close to constant relative to E$_{\text{I}}$ and E$_{\text{II}}$. For D=150 nm, (G= 30 nm), the number of T$_{\text{I}}$ and T$_{\text{II}}$ vertices are similar, which is expected in the compensated regime where all ice-rule obeying vertex configuration are equal in energy and therefore degenerate.

The spatial distribution of the vertices is shown in Fig. \ref{fig:vpopulation}c. An average domain size of 3.2 T$_{\text{I}}$ vertices is obtained from the analysis for D=0, with the largest domain composed of 37 T$_{\text{I}}$ vertices. With the discs present, the domain size of T$_{\text{I}}$ vertices is found to decrease with increasing disc diameter, reaching 1.4 for D=180 nm. At this diameter, the T$_{\text{II}}$ vertices are predominant, with an average domain size of 11.7 vertices and the largest domain observed consisting of 191 vertices. The magnetic correlation between the Ising mesospins decays rapidly, favouring the use of analysis tools which are not tied to pre-identified correlations. For example, direct entropy density estimates, as suggested by Lammert {\it et al.}\cite{Lammert2010}, can be used to obtain the upper bound for the real entropy in the arrays. 

\begin{figure*}[h!]
\includegraphics[width=0.95\textwidth]{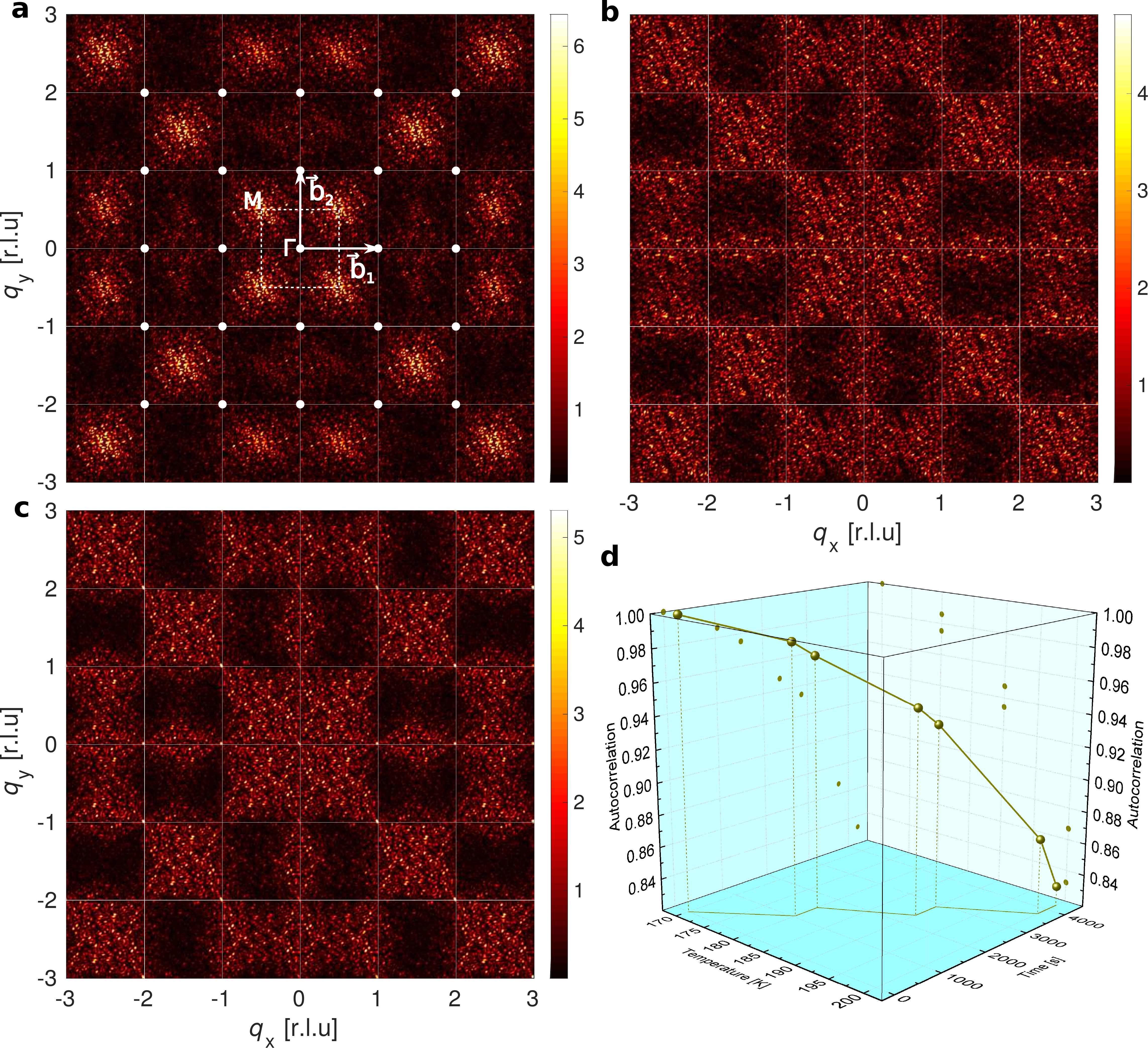}
\caption{\label{fig:ssf}\textbf{Magnetic spin structure factor and autocorrelation.} \textbf{a}, The magnetic spin structure factor for the $\alpha$=660 nm array,  D=0 nm and \textbf{b}, the magnetic spin structure factor for D=150 nm, both frozen-in spin states. The color scale indicates the intensity for every point (\textit{$q_x$},\textit{$q_y$}), while b$_1$ and b$_2$ in \textbf{a} are the reciprocal lattice vectors constructed from the real space lattice vectors in Fig. \ref{fig:intro}b. The first Brillouin zone, dotted line, is shown together with a portion of the reciprocal lattice, white dots. The $\Gamma$ point of the first Brillouin zone is also marked. \textbf{c} Time-temperature average of 7 magnetic spin structure factors exploring the spin-liquid manifold. \textbf{d} Autocorrelation of the spin states as a function of time and temperature. The autocorrelation reveals how far from the original spin state the array has evolved, an indication of how much of the spin liquid manifold has been explored.}
\end{figure*}

An even more comprehensive way to study the emergent magnetic order is the computation of the magnetic spin structure factor\cite{Canals_2016dc, Perrin2016}. Here we utilize the real space lattice and related vectors defined and presented in Fig. \ref{fig:intro}b, when calculating the magnetic spin structure factor from the real space PEEM-XMCD images. The result for $\alpha=$ 660 nm D=0 nm, as well as the reciprocal lattice, related vectors and high symmetry points are presented in Fig. \ref{fig:ssf}a. The position of the Bragg peaks at the M points of the first Brillouin zone ($[\pm \frac{1}{2},\pm \frac{1}{2}]$ reciprocal lattice units, r.l.u., or [$\frac{1}{2}$b$_1$,$\frac{1}{2}$b$_2$])  stems from domains of T$_{\text{I}}$ vertices, resulting in a magnetic structure with a periodicity twice as large as that of the lattice. The width of the peaks arises from the abundance of defects in the form of T$_{\text{III}}$ and T$_{\text{IV}}$ vertices, resulting in short correlation lengths. The thermalized system is free from kinetic constraints imposed by external fields, as e.g. observed in athermal systems \cite{Perrin2016}, yielding highly symmetric Bragg peaks. When the diameter of the discs is increased, the Bragg peaks diminish and diffuse scattering becomes more prominent (see Supplementary Fig. 4 for full data set). At D=150 nm (G=30 nm) which corresponds to the nearly compensated array, the signal is diffuse yet structured as seen in Fig. \ref{fig:ssf}b.
This map resembles the characteristic intensity distribution for a square-ice model spin liquid, associated to an emergent Coulomb phase with slow decaying spin correlations \cite{Henley_2005_PRB, Henley_2010_Coulomb}. In such cases, the spin structure factor exhibits characteristic intensity features at specific reciprocal lattice points, so called `pinch points'\cite{Fennell_2007_NatPhys}, appearing at [$\pm 1, \pm 1$] r.l.u, where the intensity exhibits a singularity\cite{Perrin2016}. The intensities at the pinch points of Fig. \ref{fig:ssf}b exhibit a weak divergence due to finite size effects and the amount of excitations in the arrays. Expressing the intensity distribution around a pinch point in polar coordinates of \{q,$\theta$\} we can identify a clear dependence on $\theta$, but not on q, in contrast with the expected intensity distribution for an ideal paramagnet, which is independent of both $\theta$ and q\cite{Frustrated_magnetism}. 


Direct observation of the magnetic microstates enables investigations of exotic magnetic phases, such as the spin liquid state \cite{Nisoli_2016_comment}. Furthermore, the rate of transformation in the spin liquid, for a given temperature and time interval, can be directly determined using the Edwards-Anderson order parameter\cite{Edwards_Anderson}. In order to probe the thermal dynamics in the spin liquid manifold, we use a heating protocol as described in Methods. Upon heating, spins will start to reverse, changing the overall magnetic structure. The time and temperature averaged spin structure factor for the nearly compensated array, computed from seven different time-temperature steps, is shown in Fig. \ref{fig:ssf}c.  Fig. \ref{fig:ssf}d illustrates the changes in the Edwards-Anderson order parameter\cite{Edwards_Anderson} which we use to determine how far from the original spin configuration the array has evolved. It has the form of an autocorrelation between the measurement at t=t$_0$ and every subsequent time-temperature step (see Methods for further details) as shown in Fig. \ref{fig:ssf}d. Between t=t$_0$ and t=$\text{t}_{final}$ close to 20 \% of all spins have reversed, while the array still remains in the spin liquid-like state. The evolution of the system presented in Fig. \ref{fig:ssf}d depicts changes in the spin liquid manifold, differing significantly from other dynamics studies targeting thermal relaxation processes\cite{Farhan_2013_NatPhys, Farhan2013, Andersson_2016_SciRep}. Here the magnetic structure does not relax, instead the vertex populations and domain sizes remain constant. The activities are similar for the T$_{\text{I}}$, T$_{\text{II}}$, and T$_{\text{III}}$ vertices, while T$_{\text{IV}}$ vertices are comparably more active (see Supplementary Fig. 11). This approach enables the investigation of mesospin dynamics in magnetic frustrated materials, thereby allowing to shed light on the related glassy dynamics and monitor the evolution of liquid-like spin configurations in time and temperature\cite{Colloids_dynamics_PRX_2017}.

\begin{figure*}
\includegraphics[width=0.95\textwidth]{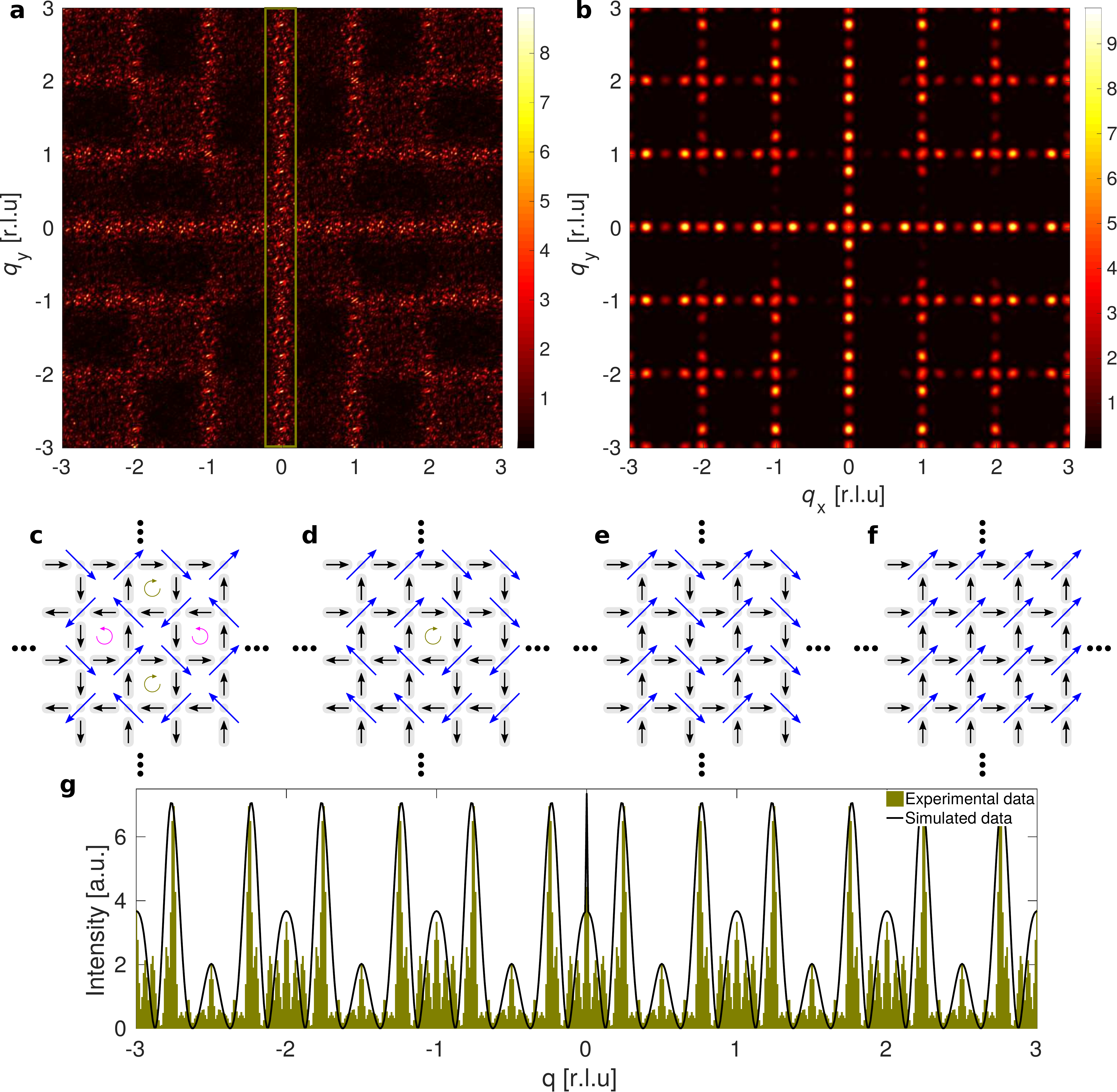}
\caption{\label{fig:spinstates}\textbf{Emergent flux lattice ordering}. \textbf{a}, Magnetic spin structure factor of the $\alpha$=660 D=180 array. \textbf{b}, Summation of the magnetic structure factor for spin state \textbf{c} along with spin states \textbf{d}, \textbf{e} and its $\pi$/2 rotation, and \textbf{f} and its $\pi$/2 rotation. The magnetic flux from T$_{\text{II}}$ vertices is drawn with blue arrows. \textbf{g} cut along q$_x$=0 for both the experimental and simulated maps. The intensity of the peaks in the simulated map are scaled to match the experimental data.} 
\end{figure*}

At a disc diameter of D=180 nm, the T$_{\text{II}}$ population increases to almost 60 \%, while the T$_{\text{I}}$ population decreases below 30 \% with vertex energies E$_{\text{II}}<\text{E}_{\text{I}}<\text{E}_{\text{III}}<\text{E}_{\text{IV}}$. The resulting spin structure factor looks completely different as illustrated in Fig. \ref{fig:spinstates}a (see also the PhD thesis of Y. Perrin for a discusion concerning an athermal system\cite{Perrin2016_phd}). When considering the effective dipole moment associated with the T$_{\text{II}}$ vertices, it becomes clear that their abundance can give rise to an emergent flux lattice on the next length scale, dictating the magnetic order of the spin system. In more detail, we experimentally identify four different types of flux lattices (see Supplementary Fig. 10): An emergent T$_{\text{I}}$-like tiling of T$_{\text{II}}$ vertices (see Fig. \ref{fig:spinstates}c), vortex/anti-vortex pairs (see Fig. \ref{fig:spinstates}d), and  ferromagnetic states, forming both a non-collinear herringbone structure (Fig. \ref{fig:spinstates}e) or collinear ferromagnetic domains (Fig. \ref{fig:spinstates}f).  
The flux lattice with the lowest energy, as determined by micro-magnetic calculations, is the emergent T$_{\text{I}}$-like tilling, whereas the vortex structure has only somewhat higher energy. Both the ferromagnetic states invoke the presence of a net moment, implying a higher energy cost and being more unfavourable, as compared to the flux closure states.
The energy differences $\Delta$E between the ground state and the emergent states, expressed as $\Delta$E/k$_B$, are 5 K, 58 K, and 123 K for the vortex-antivortex state, the herringbone state, and the polarized state respectively (calculated using micro-magnetic simulations). 
The energy difference between the T$_{\text{I}}$-tiling and the collinear flux state is therefore almost 25 times larger than the difference between the T$_{\text{I}}$-tiling and the vortex-antivortex state. In order to get an estimate of the magnetic ordering of the array we have calculated the spin structure factor maps for the configurations depicted in Fig. \ref{fig:spinstates}c-f and summed all four results into one map in Fig. \ref{fig:spinstates}b (see Methods for details). As seen in Fig. \ref{fig:spinstates}b, there is a strong resemblance between the experimental and the simulated results. 
To further elaborate on this we make a cut along $q_x$=0 for both the experimental and the calculated result shown in Fig. \ref{fig:spinstates}a and b. The weighted sum obtained from the four textures described in Fig. \ref{fig:spinstates}c-f overlap to a great extent with the experimental data. With long-range interactions present\cite{Rougemaille2011}, the state with the lowest energy in the emergent flux lattice is, as already mentioned, the two fold degenerate T$_{\text{I}}$-tiling. This state is from a symmetry perspective, identical to the ground state in regular SASI, albeit at a different length scale. At finite temperatures we observe the traces of competing states, with small energy differences, as well as frozen-in higher energy states. In the scenario of an array with even more dominant populations of T$_{\text{II}}$ vertices, which implies even larger energy gaps between T$_{\text{II}}$ vertices and other vertex types, the T$_{\text{II}}$ abundance would make the features we observe even more pronounced. This could provide a pathway towards examining systems where order is dominated by the emergent flux lattices.

We have presented a generic solution to continuously alter the effective coupling between mesoscopically-sized islands of a ferromagnetic material in a fully planar geometry. This approach can be utilized to engineer the energy landscape of two-dimensional nanomagnetic systems in a completely new way, employing nanomagnetic objects of distinctively different mesospin dimensionality. In our example using nanosized magnetic discs, we tailor the energy landscape of arrays, recovering the degeneracy in SASI and also promoting emergent magnetic order of the Ising mesospins. This approach opens thereby new routes for investigations of ordered and frustrated artificial systems. Here we have only focused on the final state of one of the subsystems in the sample (elongated Ising-like islands), treating the discs as interaction modifiers. One can also envisage structures where the situation is reversed and the collective magnetic structure of the discs dominates the ordering, leaving the Ising mesospins in the role of the modifier. This synergy and cooperative behaviour therefore provides a route for designing new types of magnetic metamaterials with rich magnetic phase diagrams and thermodynamics. 
The calculated spin structure factors, obtained from the real space microscopy results, demonstrate the plausibility of using magnetic scattering \cite{Morgan_2012_XRMS, Sendetskyi_2016_XRMS}, providing new insights on emergent mesoscale magnetic structures driven by collective dynamics\cite{Nisoli_2017_NatPhys}.

\subsection{Acknowledgments}
The authors would like to thank S.T. Bramwell and P.C.W. Holdsworth for valuable discussions. The authors acknowledge support from the Knut and Alice Wallenberg Foundation, the Swedish Research Council and the Swedish Foundation for International Cooperation in Research and Higher Education. The patterning was performed at the Center for Functional Nanomaterials, Brookhaven National Laboratory, supported by the U.S. Department of Energy, Office of Basic Energy Sciences, under Contract No. DE-SC0012704. This research used resources of the Advanced Light Source, which is a DOE Office of Science User Facility under contract no. DE-AC02-05CH11231. This work is part of a project that has received funding from the European Union's Horizon 2020 research and innovation programme under grant agreement No 713171. U.B.A. acknowledges funding from the Icelandic Research Fund grants nr. 141518 and 152483.

\subsection{Author contributions}
H.S. and A.S. fabricated the sample. E.\"{O}., H.S., U.B.A., and V.K. performed the PEEM-XMCD experiments. E.\"{O}., I.A.C., H. S., V.K., and B.H. analyzed the data and contributed to theory development. E.\"{O}., I.A.C., V.K., and B.H. wrote the manuscript. All authors discussed the results and commented on the manuscript.

\bibliographystyle{natphys}
\bibliography{mylib}

\blue{\section{Methods}} \label{sec:methods}

\subsection{Sample manufacturing}
The arrays were patterned from $\delta$-doped Pd(Fe) \cite{Papaioannou2010} thin films grown on MgO substrates with a 1,5 nm thick V seeding layer using a UHV sputter system. $\delta$-doped Pd(Fe) is a magnetic trilayer system, in our case consisting of Pd(40 nm)/Fe(2.0 monolayers)/Pd(2 nm). The magnetic nano-structures were produced by post-patterning the Pd(Fe) $\delta$-doped thin films using e-beam lithography at the Center for Functional Nanomaterials (CFN), Brookhaven National Laboratory in Upton New York. The physical dimension of each array is 200 $\times$ 200 $\mu$m$^2$ and were all patterned on the same substrate ensuring near-identical conditions for all arrays during the measurements.

\subsection{PEEM-XMCD}

The Photoemission Electron Microscopy (PEEM) measurements employing the X-ray Magnetic Circular Dichroism (XMCD) technique were carried out at the 11.0.1 PEEM3 beamline at the Advanced Light Source, CA, USA. The imaging of the frozen states was performed at a temperature of 120 K using the L$_3$ edge of Fe ($708.6~\rm{eV}$). For each array multiple XMCD images where acquired and merged together revealing the state of several thousands of islands. Due to the size of the discs it is difficult to observe their magnetic orientation, however some of them can be observed in the PEEM-XMCD images (see Supplementary Fig. 9). At 120 K all mesospins are frozen with an average fluctuation rate lower than the time-scale of the whole experiment, as such no mesospin fluctuations were observed at 120 K. This also have the implication that changing the acquisition protocol in this frozen regime do not affect the observed state.

\subsection{Magnetic spin structure factor}
The magnetic spin structure factor is defined analogous with neutron scattering experiments where spin correlations perpendicular to the scattering vector is measured.
We start by defining a perpendicular spin component $\mathbf{S}^{\perp}$ of spin $\mathbf{S}$:
\begin{equation}
  \mathbf{S}^{\perp}=\mathbf{S}-(\hat{\mathbf{q}}\cdot \mathbf{S})\hat{\mathbf{q}}
\end{equation}
where $\hat{\mathbf{q}}$ is the unit scattering vector:
\begin{equation*}
\hat{\mathbf{q}}=\frac{\mathbf{q}}{\|\mathbf{q}\|}
\end{equation*}
For every $\mathbf{q}=(q_x,q_y)$ the intensity I is given by:
\begin{equation}
 I(\mathbf{q})=\frac{1}{N}\sum_{(i,j=1)}^{N}\mathbf{S}^{\perp}_i\cdot \mathbf{S}^{\perp}_j exp(i\mathbf{q}\cdot(\mathbf{r}_i-\mathbf{r}_j))
\end{equation}
Which we can write as:
\begin{equation}
 I(\mathbf{q})=\frac{1}{N} \left(\sum_{i=1}^{N}\mathbf{S}^{\perp}_i exp(i\mathbf{q}\mathbf{r}_i)\right) \cdot \left(\sum_{j=1}^{N}\mathbf{S}^{\perp}_j exp(-i\mathbf{q}\mathbf{r}_j)\right)
\end{equation}
Expanding yields:
\begin{equation}
 I(\mathbf{q})=\frac{1}{N} \left(\sum_{i=1}^{N}\mathbf{S}^{\perp}_i cos(\mathbf{q}\cdot\mathbf{r}_i)+i\sum_{i=1}^{N}\mathbf{S}^{\perp}_i sin(\mathbf{q}\cdot\mathbf{r}_i)\right)\cdot\left(\sum_{j=1}^{N}\mathbf{S}^{\perp}_j cos(\mathbf{q}\cdot\mathbf{r}_j)-i\sum_{j=1}^{N}\mathbf{S}^{\perp}_j sin(\mathbf{q}\cdot\mathbf{r}_j)\right)
\end{equation}
Recognizing that i and j sums up over the same spins and defining $\mathbf{A}=\sum_{i=1}^{N}\mathbf{S}^{\perp}_i cos(\mathbf{q}\cdot\mathbf{r}_i)$ and $\mathbf{B}=\sum_{i=1}^{N}\mathbf{S}^{\perp}_i sin(\mathbf{q}\cdot\mathbf{r}_i)$ we can simplify the equation such that:
\begin{equation}
 I(\mathbf{q})=\frac{1}{N}\left(\mathbf{A}+i\mathbf{B}\right)\cdot\left(\mathbf{A}-i\mathbf{B}\right)=\frac{1}{N}\left(\mathbf{A}^2+\mathbf{B}^2\right)
\end{equation}
I is now a real quantity which we calculate for the interval $(q_x,q_y)$=[-3,-3]-[3,3] r.l.u. in 601x601 steps.

\subsection{Heating protocol}
The sample was cooled from its paramagnetic state to 170 K where the spin flip time is in the order of hours. At this temperature the $\text{t}_{0}$ spin state was recorded. The temperature was subsequently raised in steps of 10 K up to 200 K with two measurement points recorded at each temperature. The nominal acquisition time was kept the same for all measurements with the shortest acquisition time being 470 s and the longest 483 s. The starting time t for all measurements relative to $\text{t}_0$ were 1047 s, 1531 s, 2371 s,2864 s, 3753 s, and 4242 s.

\subsection{Autocorrelation}
The autocorrelation is calculated in a way so that any change in the spin system at $\text{t}>\text{t}_0$ is tracked cumulatively towards $\text{t}_{final}$. For every given time-temperature (t,T) step the autocorrelation is given by $\text{Q}(\text{t},\text{T})=\frac{1}{N}\sum_{\text{j}=1}^{N}\mathbf{S}_{\text{j},\text{t}_0,\text{T}_i}\cdot\mathbf{S}_{\text{j},\text{t},\text{T}}$ where $\text{t}_0$ is the inital time, T$_i$ is the initial temperature, and N is the number of islands with an assignable magnetic vector. Only islands visible in both time-temperature steps are taken into account (see Supplementary Fig. 8).

\subsection{Composite spin structure map}
In order to obtain insight on the overall spin structure of the $\alpha$=660, D=180 array, see Fig. \ref{fig:spinstates}a, we calculated the individual spin structure factor of the states illustrated in \ref{fig:spinstates} c-f in the following way. We used an array containing 144 islands, with 64 vertices, all T$_{\text{II}}$. Concerning the structure illustrated in Fig. \ref{fig:spinstates}d this implies that the spin structure factor is calculated from 5 vortices and 4 antivortices. Each of the spin structure factors are scaled in intensity (1/20, 1/2, 1/9, 1/20) in order to match the experimental data illustrated in Fig. \ref{fig:spinstates}f. The weighted sum of the four spin structure factor maps is shown in Fig. \ref{fig:spinstates}b
The data points are binned in series of three, using a moving average, in the bar diagram in Fig. \ref{fig:spinstates}g.

\subsection{Micromagnetic simulations}
The micro-magnetic simulations were performed using the GPU-accelerated M{\scriptsize U}M{\scriptsize AX}3 software\cite{Vansteenkiste2014}. The calculations are all 0 K calculations with a saturation magnetization of Ms=560320 A/m\cite{Vassilios2014} and an exchange stiffness of 6.5$\times10^{-12}$. The thickness of the magnetic layers was assumed to be 1 nm. The calculation of the energies for the states illustrated in Fig. \ref{fig:spinstates} where performed using 32 islands and 16 discs, using periodic boundaries. Initially, the magnetic order was pre-defined in all elements. The system was thereafter relaxed, a process where M{\scriptsize U}M{\scriptsize AX}3 minimizes the energy, allowing for divergence of the magnetization within the elements.

\end{document}


\title{The importance of the weak: Interaction modifiers in artificial spin ices}

\author{Erik \"{O}stman}
\email{erik.ostman@physics.uu.se}
\affiliation{%
 Department of Physics and Astronomy, Uppsala University, Box 516, SE-75120, Uppsala, Sweden
}%
\author{Henry Stopfel}%
\affiliation{%
 Department of Physics and Astronomy, Uppsala University, Box 516, SE-75120, Uppsala, Sweden
}%
\author{Ioan-Augustin Chioar}%
\affiliation{%
 Department of Physics and Astronomy, Uppsala University, Box 516, SE-75120, Uppsala, Sweden
}%
\author{Unnar B. Arnalds}%
\affiliation{%
 Science Institute, University of Iceland, Dunhaga 3, 107 Reykjavik, Iceland
}%
\author{Aaron Stein}%
\affiliation{%
 Center for Functional Nanomaterials, Brookhaven National Laboratory, Upton, New York 11973, USA
}%
\author{Vassilios Kapaklis}%
\affiliation{%
 Department of Physics and Astronomy, Uppsala University, Box 516, SE-75120, Uppsala, Sweden
}%
\author{Bj\"{o}rgvin Hj\"{o}rvarsson}%
\affiliation{%
 Department of Physics and Astronomy, Uppsala University, Box 516, SE-75120, Uppsala, Sweden
}%

\date{\today}
\maketitle

\begin{figure}[h]
\includegraphics[width=0.36\columnwidth]{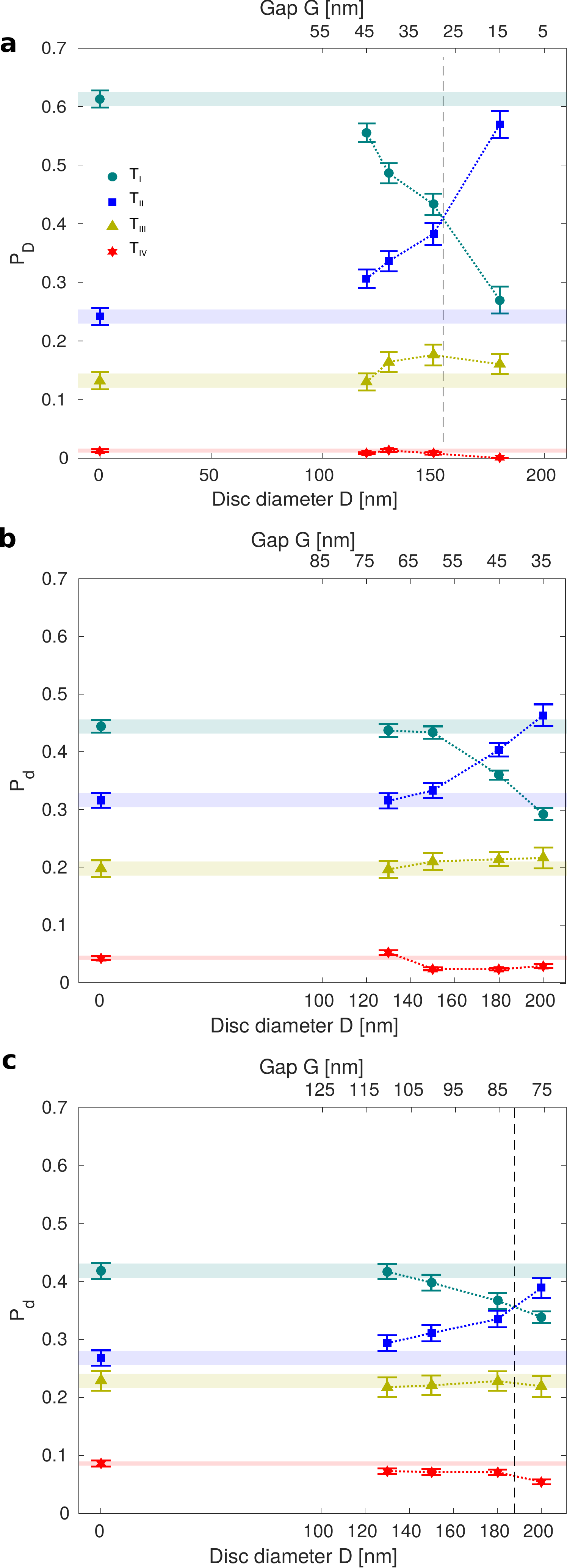}
\caption{\label{fig:a1} Degeneracy corrected and normalized vertex populations for $\alpha=660$ nm \textbf{a}, $\alpha=720$ nm \textbf{b}, and $\alpha=800$ nm \textbf{c}. Vertical dashed lines indicate the intersection between $\text{T}_\text{I}$ and $\text{T}_\text{II}$ vertices drawing straight lines betweeen the data points. The horizontal coloured stripes refer to the populations for D=0 and serve as a guide to the eye along with the dotted lines, highlighting the vertex population changes while varying D. All three groups with different lattice parameters undergo a transition from a majority of $\text{T}_\text{I}$ vertices to a majority of $\text{T}_\text{II}$ vertices with an increase in disc diameter. The increase in lattice parameter decreases the coupling strength between the interacting elements evident from the increase of charges for the $\alpha=720$ nm and the $\alpha=800$ nm lattices.}
\end{figure}

\begin{figure}[h]
\includegraphics[width=0.99\columnwidth]{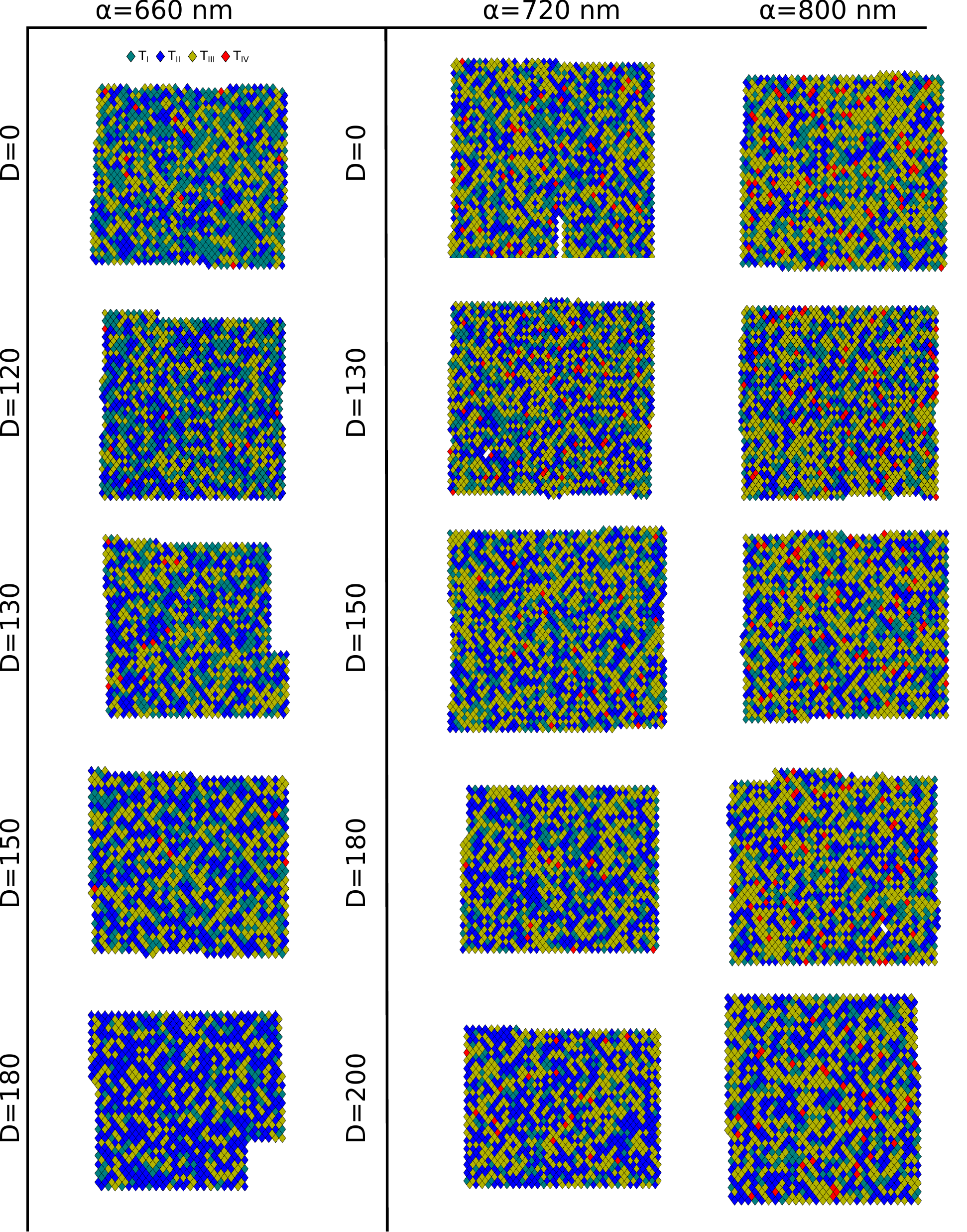}
\caption{\label{fig:a2} Vertex maps for all measured arrays. The lattice parameter, $\alpha$ is defined for each vertical line, the diameter,D, is defined for every lattice horizontally.} 
\end{figure}

\begin{figure}[h]
\includegraphics[width=0.5\columnwidth]{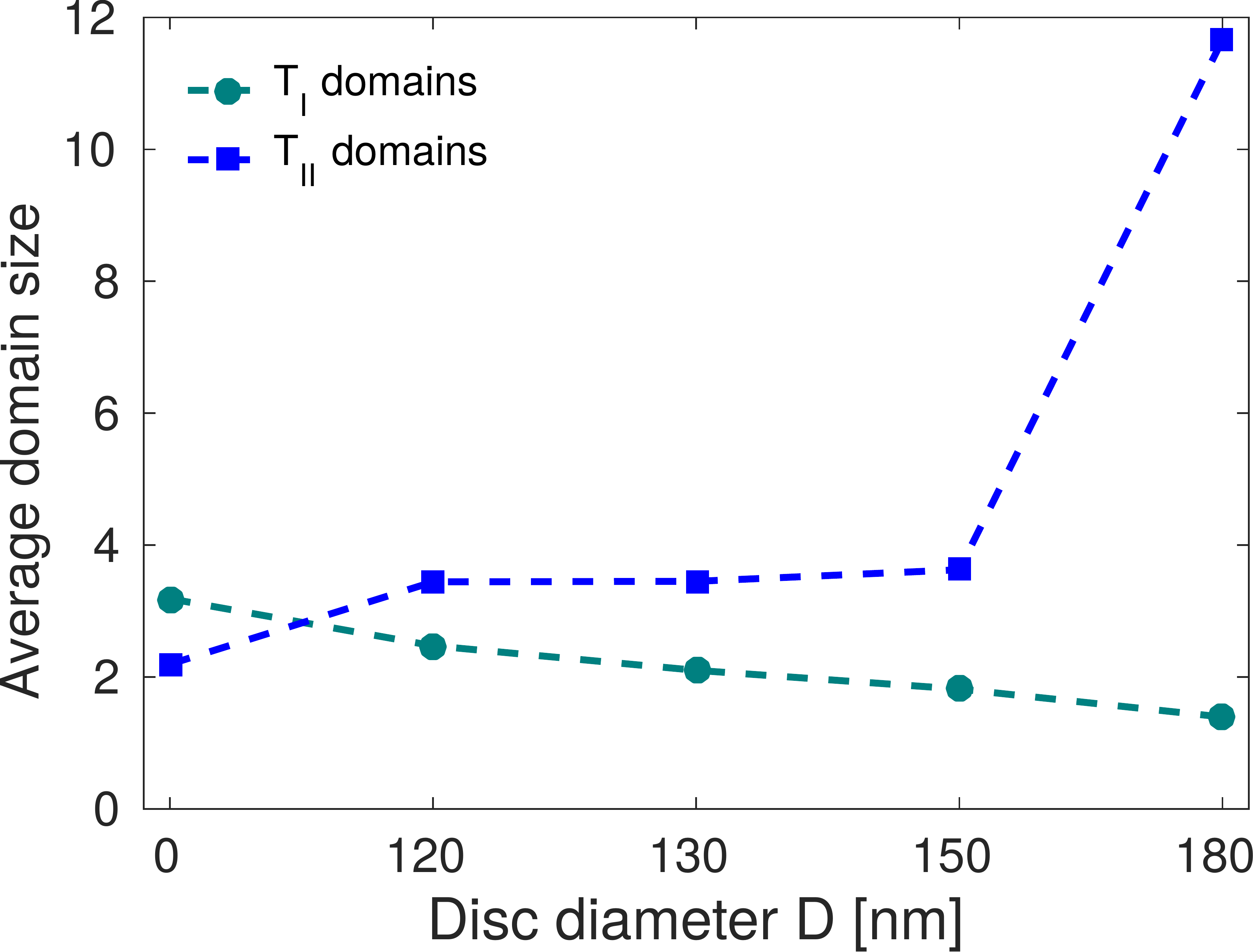}
\caption{\label{fig:a3} Average domain size evolution for $\text{T}_\text{I}$ and $\text{T}_\text{II}$ vertices with increasing disc diameter.} 
\end{figure}

\begin{figure}[h]
\includegraphics[width=0.95\columnwidth]{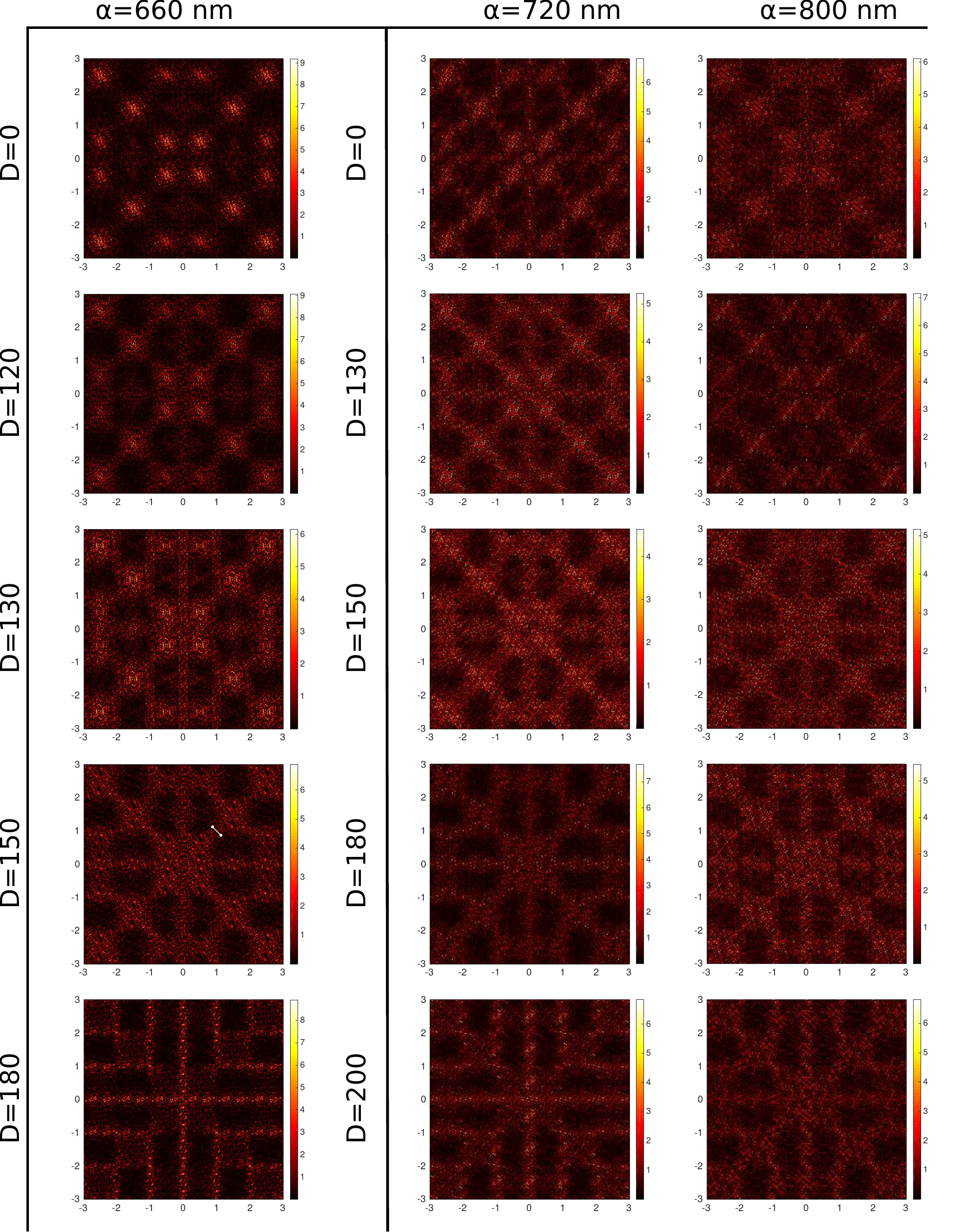}
\caption{\label{fig:a4} Magnetic spin structure factor for all meausured arrays. The white line in $\alpha=660$ nm D=150 nm illustrates the cut in reciprocal space where from the pinch point data in Supplementary Fig. 5 \textbf{a} is taken.}
\end{figure}

\begin{figure}[h]
\includegraphics[width=0.95\columnwidth]{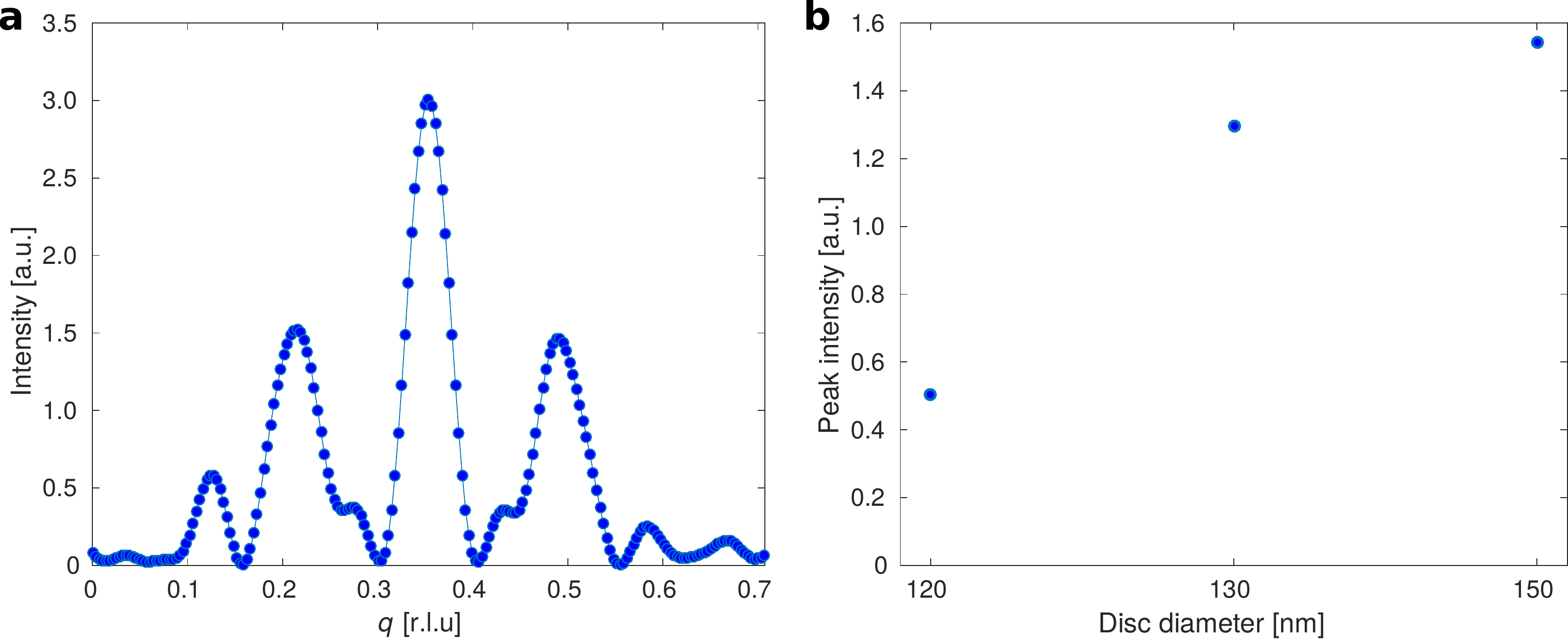}
\caption{\label{fig:a5} \textbf{a} Intensity profile of the $\alpha=660$ nm D=150 lattice going from [3/4, 5/4] to [5/4, 3/4] r.l.u passing through the pinch point at [1,1] r.l.u in reciprocal space, see Supplementary figure \ref{fig:a4}. \textbf{b} Evolution of peak intensity values at the pinch points while increasing the disc diameter. Peak values are averaged over all four positions $[\pm 1,\pm 1]$ r.l.u.}
\end{figure}

\begin{figure}[h]
\includegraphics[width=0.95\columnwidth]{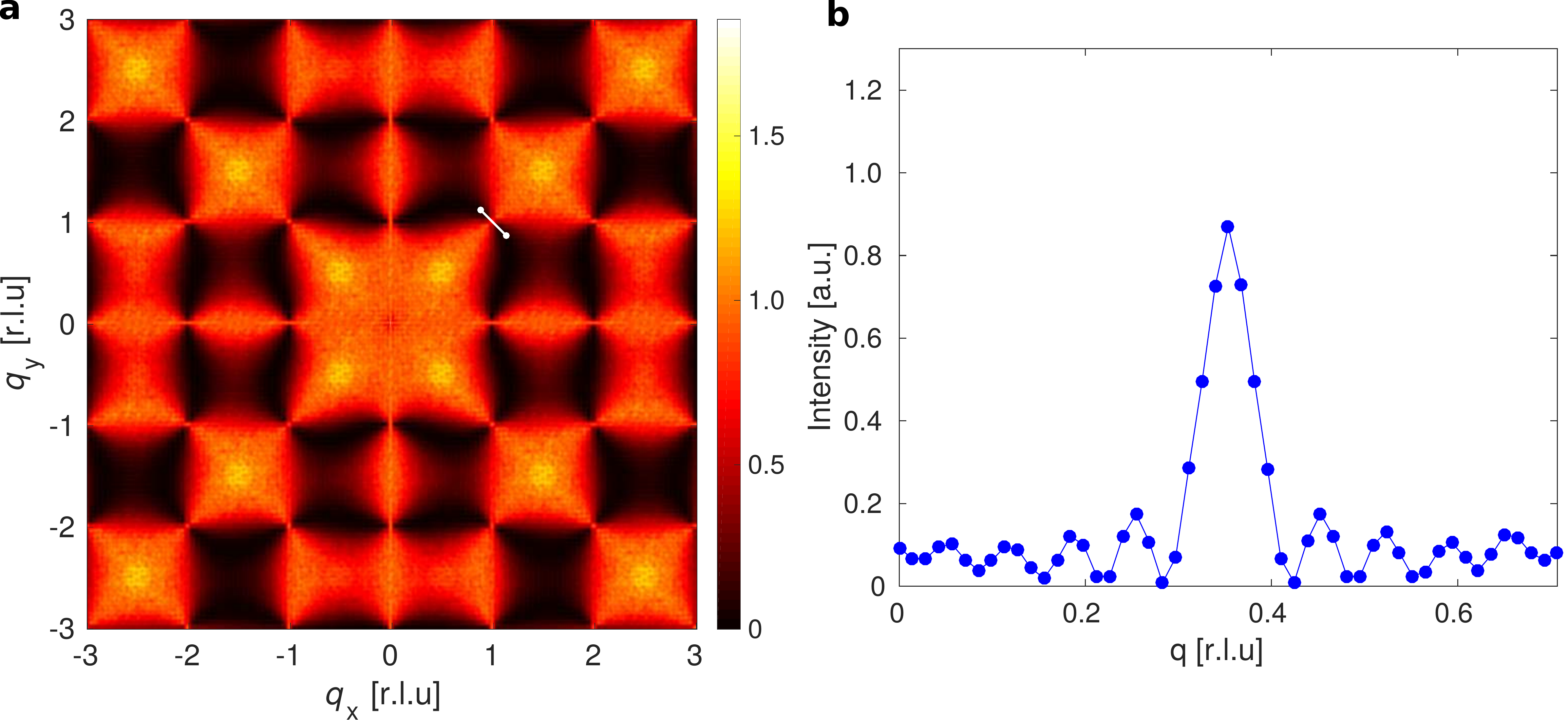}
\caption{\label{fig:a6} \textbf{a} Computed magnetic spin structure factor averaged over 800 decorrelated spin configurations satisfying the ice rule. In order to calculate a theoretical spin map for a fully compensated lattice we start with a spin configuration containing N=882 spins using periodic boundary conditions polarized diagonally with respect to the major axis of the square lattice and employ the standard loop algorithm\cite{Barkema1998,Melko_Loop_PRL_2001,Perrin2016}. To fully decorrelate the spin configurations 4N loops are flipped. For each loop the first spin is choosen at random in the lattice, the next spin is chosen from one of the six neighbors at random with the condition that the spins in the loop must align ferromagnetically after the spin flip. As soon as the loop is closed all trailing spins not contained in the loop are flipped back and a new loop is started. The loops can either be closed, contained within the lattice, or open, wrapped around the lattice. In this way all vertices will obey the ice rule and only the ice rule manifold will be explored. The calculated magnetic spin structure factor map is averaged over 800 decorrelated spin configurations. \textbf{b} Intensity profile over [3/4, 5/4] to [5/4, 3/4] r.l.u passing through the pinch point at [1,1] r.l.u., as indicated by the white line in \textbf{a}.}
\end{figure}

\begin{figure}[h]
\includegraphics[width=0.99\columnwidth]{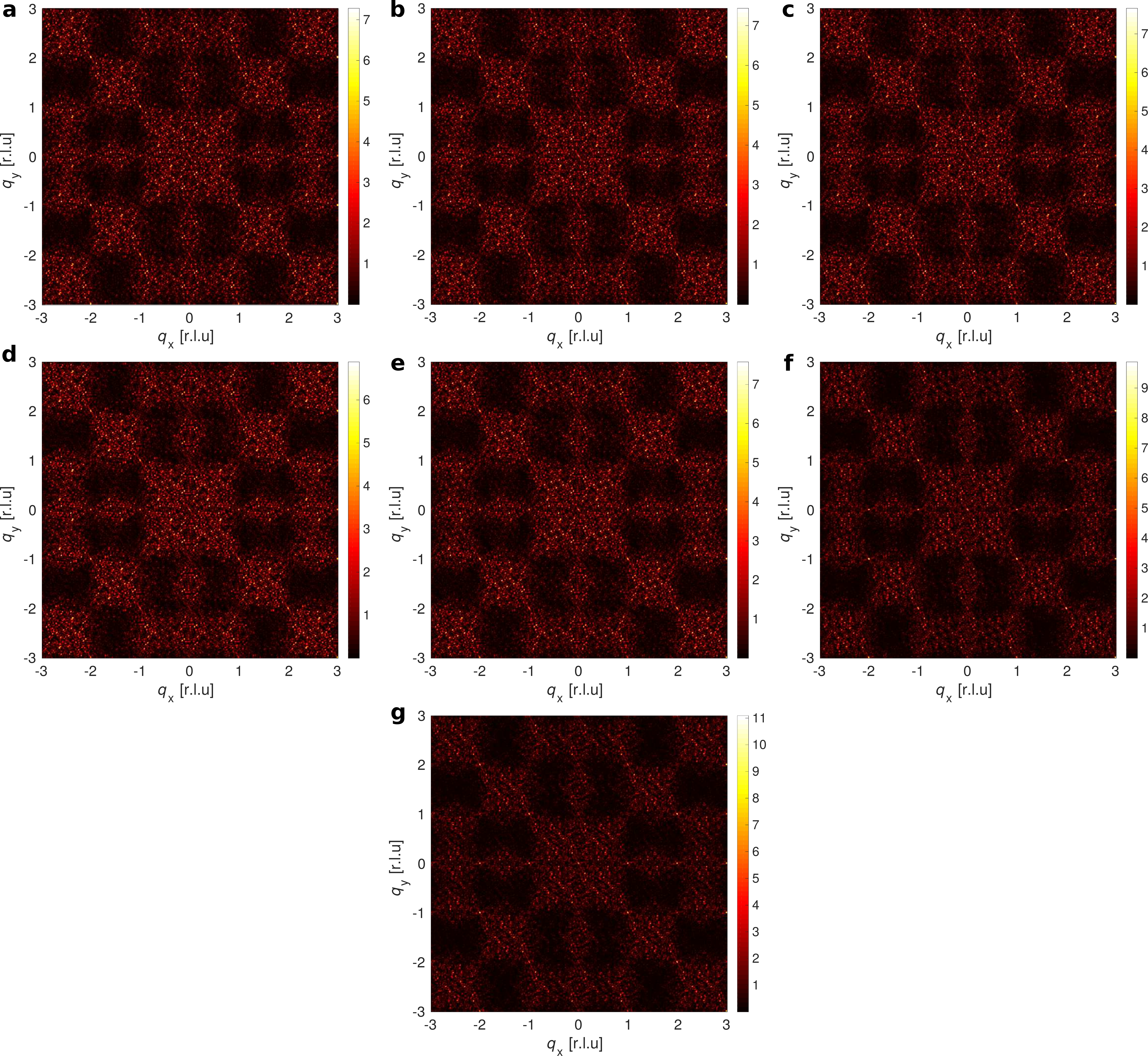}
\caption{\label{fig:a7} Computed magnetic spin structure factor for the time-temperature series. \textbf{a} corresponds to 170 K t=t$_0$, \textbf{b} 180K t=1047 s, \textbf{c} 180 K t=1531 s, \textbf{d} 190 K t=2371 s, \textbf{e} 190 K t=2864 s, \textbf{f} 200 K t=3753 s, \textbf{g} 200 K t=4242 s.}
\end{figure}

\begin{figure}[h]
\includegraphics[width=0.5\columnwidth]{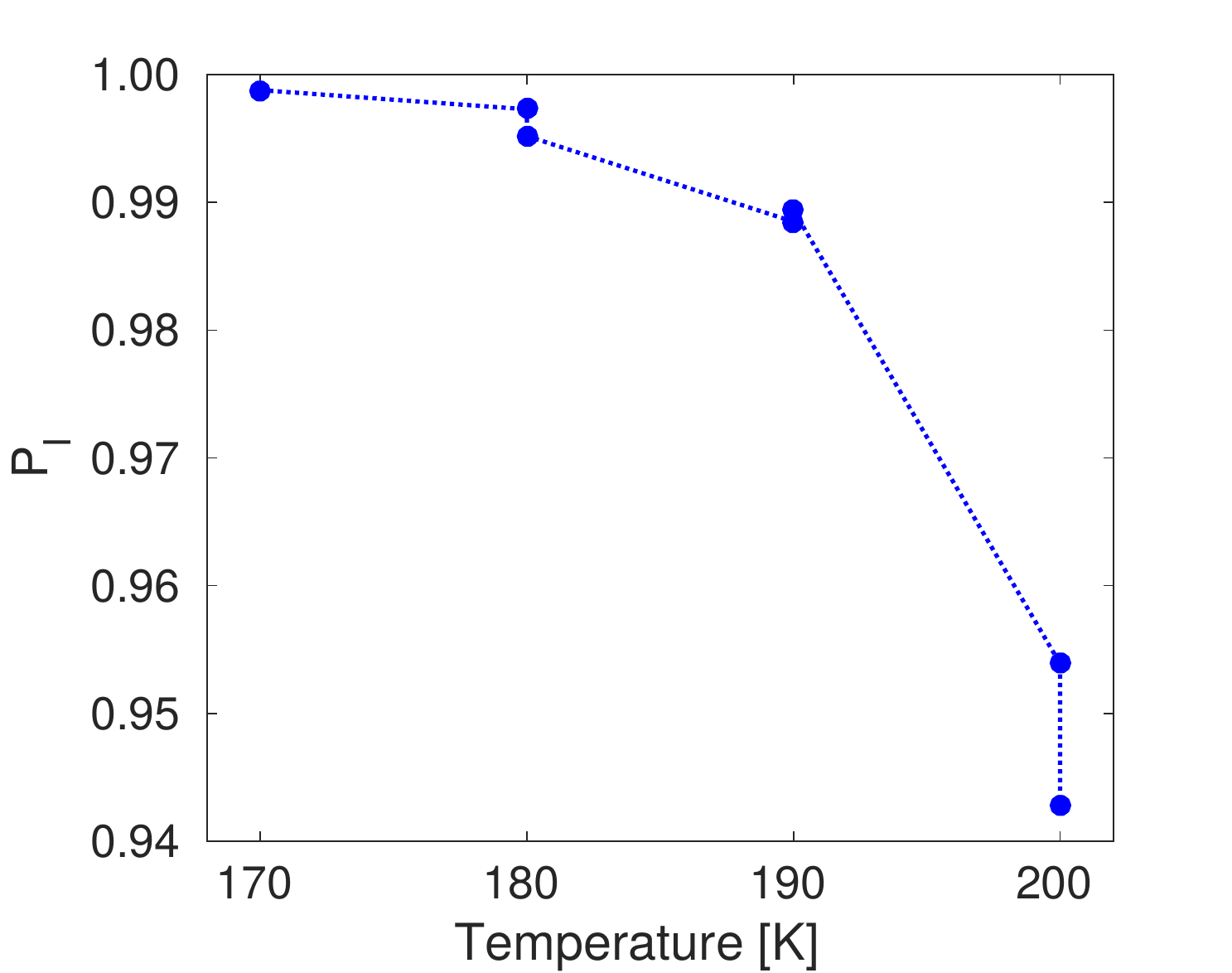}
\caption{\label{fig:a8} Population of islands with a distinct magnetization direction, P$_\text{I}$, for each time-temperature step. An island spending equal, or near equal, amount of time in each of it's two possible directions during the measurement will not show any magnetic contrast.}
\end{figure}

\begin{figure}[h]
\includegraphics[width=0.8\columnwidth]{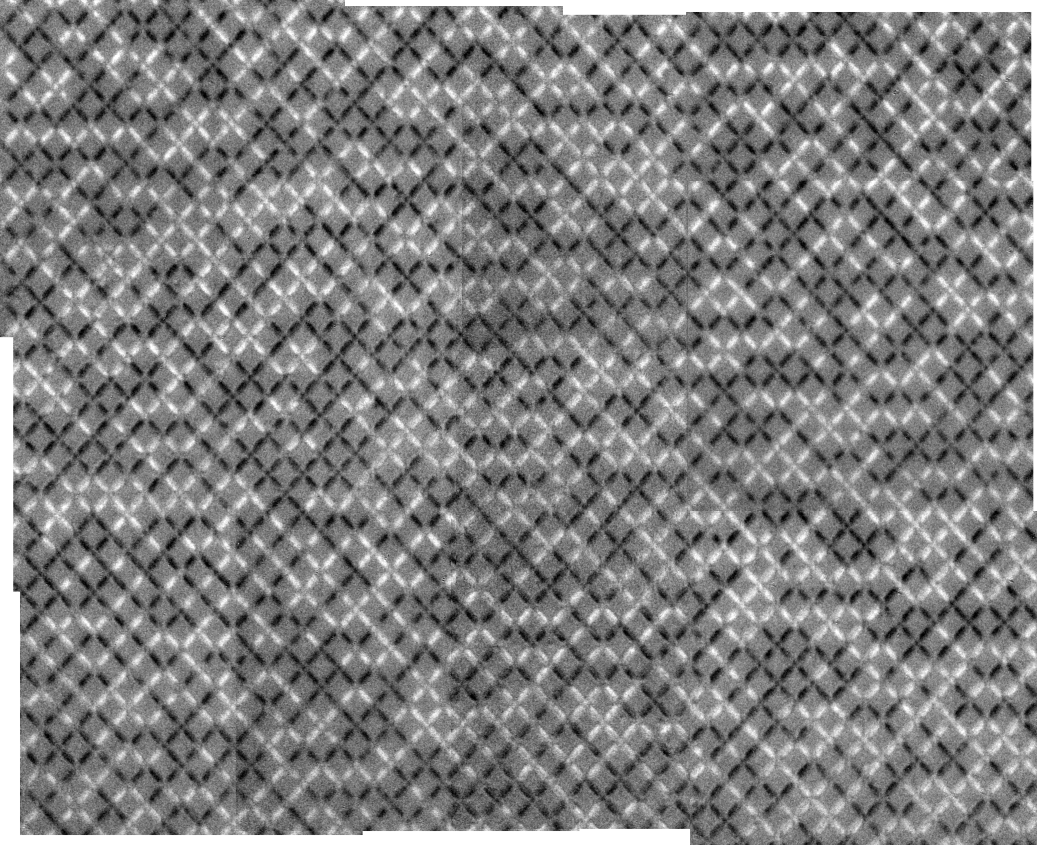}
\caption{\label{fig:a9} PEEM-XMCD image of the $\alpha=660$ nm D=150 nm lattice. The islands orientation is rotated 45 degrees with respect to the image shown in Fig. 2 in the article. The image is merged from 12 individual PEEM-XMCD images. The quality of the image is representative for all lattices measured.}
\end{figure}

\begin{figure}[h!]
\includegraphics[width=0.8\columnwidth]{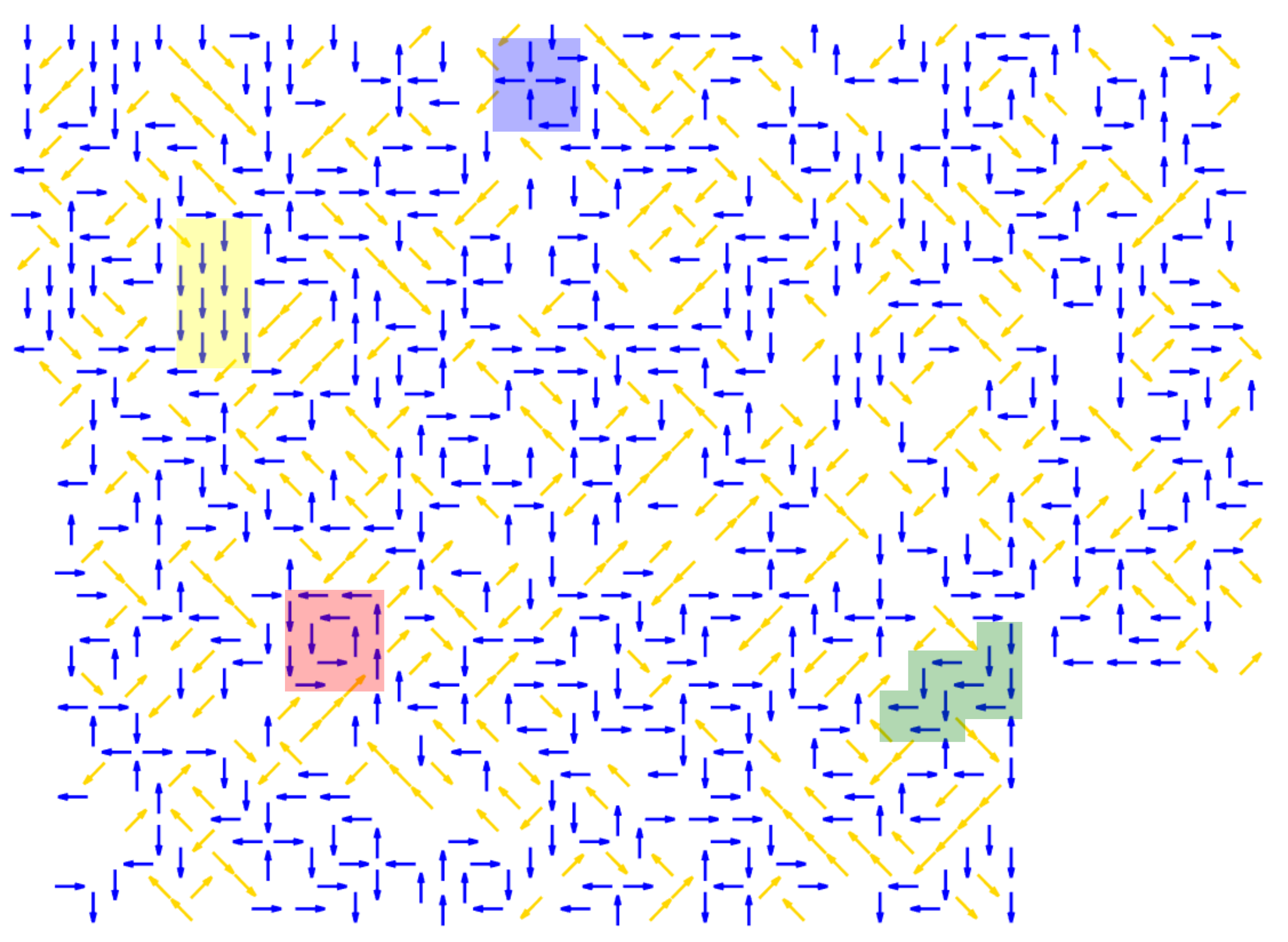}
\caption{\label{fig:a10} Magnetic flux map of the $\alpha=660$ nm D=180 nm lattice with residual flux from $\text{T}_\text{II}$, blue, and $\text{T}_\text{III}$, yellow, vertices. The map is rotated 45 degrees in comparison to the schematic maps in Fig. 4 in the article. Different $\text{T}_\text{II}$ flux ordering are marked out. Ferromagnetic order, yellow box, herringbone structure, green box, $\text{T}_\text{I}$ tiling, blue box and larger scale vortex, red box.}
\end{figure}

\begin{figure}[h]
\includegraphics[width=0.5\columnwidth]{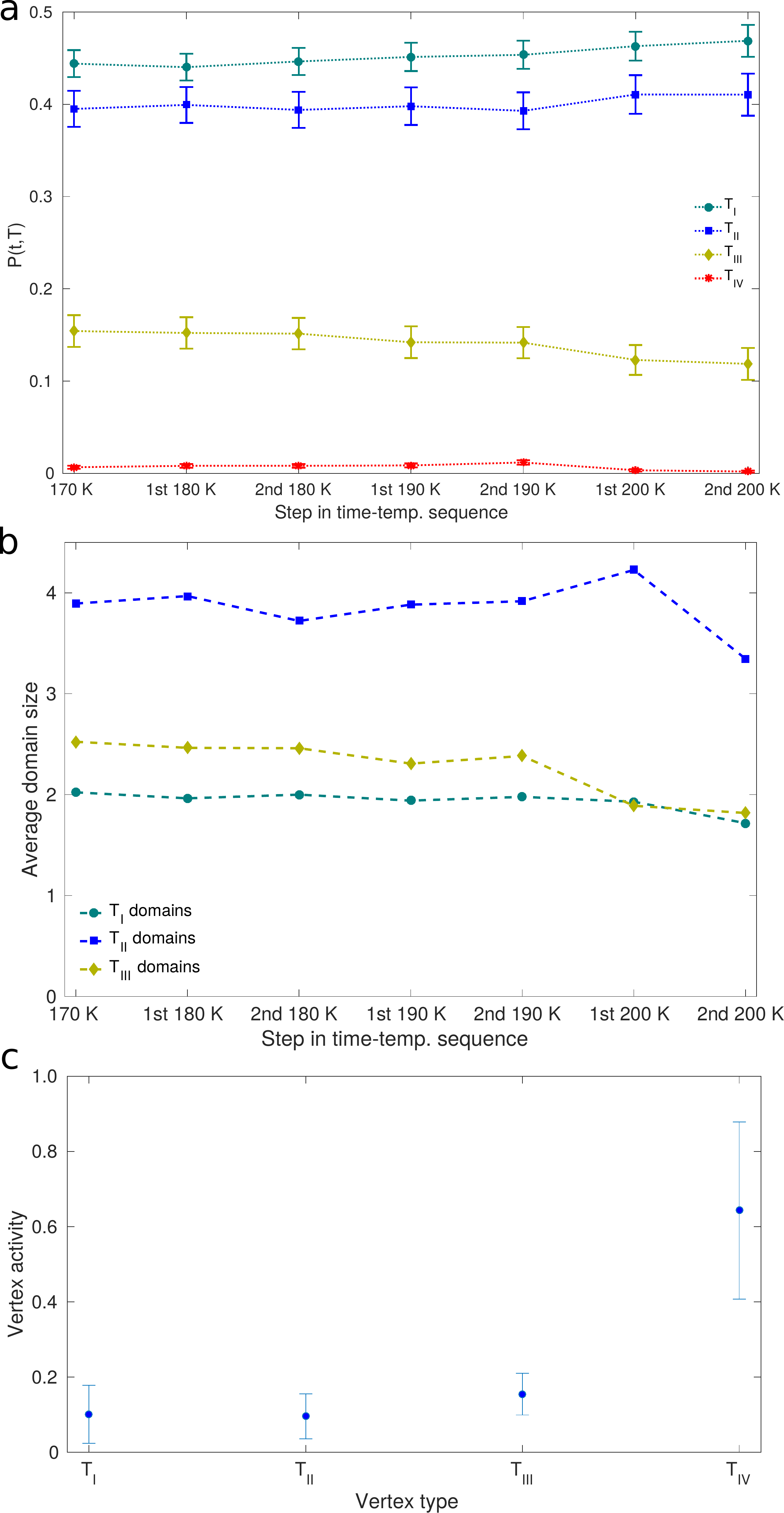}
\caption{\label{fig:a11} \textbf{a}, Evolution of the degeneracy corrected, normalized, population during the time-temperature sequence. The populations remains almost constant for the measurement interval. \textbf{b}, Evolution of domain sizes for the different vertex types during the time-temperature sequence. The slight decrease in domain sizes is attributed to the decreasing value of islands with a distinct direction of magnetization, see Supplementary Figure \ref{fig:a8}. \textbf{c}, Normalized cummulative vertex activity, counted as all vertex transitions between all time-temperature steps for the vertex types individually, normalized to each vertex type count for all time-temperature steps.}
\end{figure}

\begin{figure}[h]
\includegraphics[width=0.5\columnwidth]{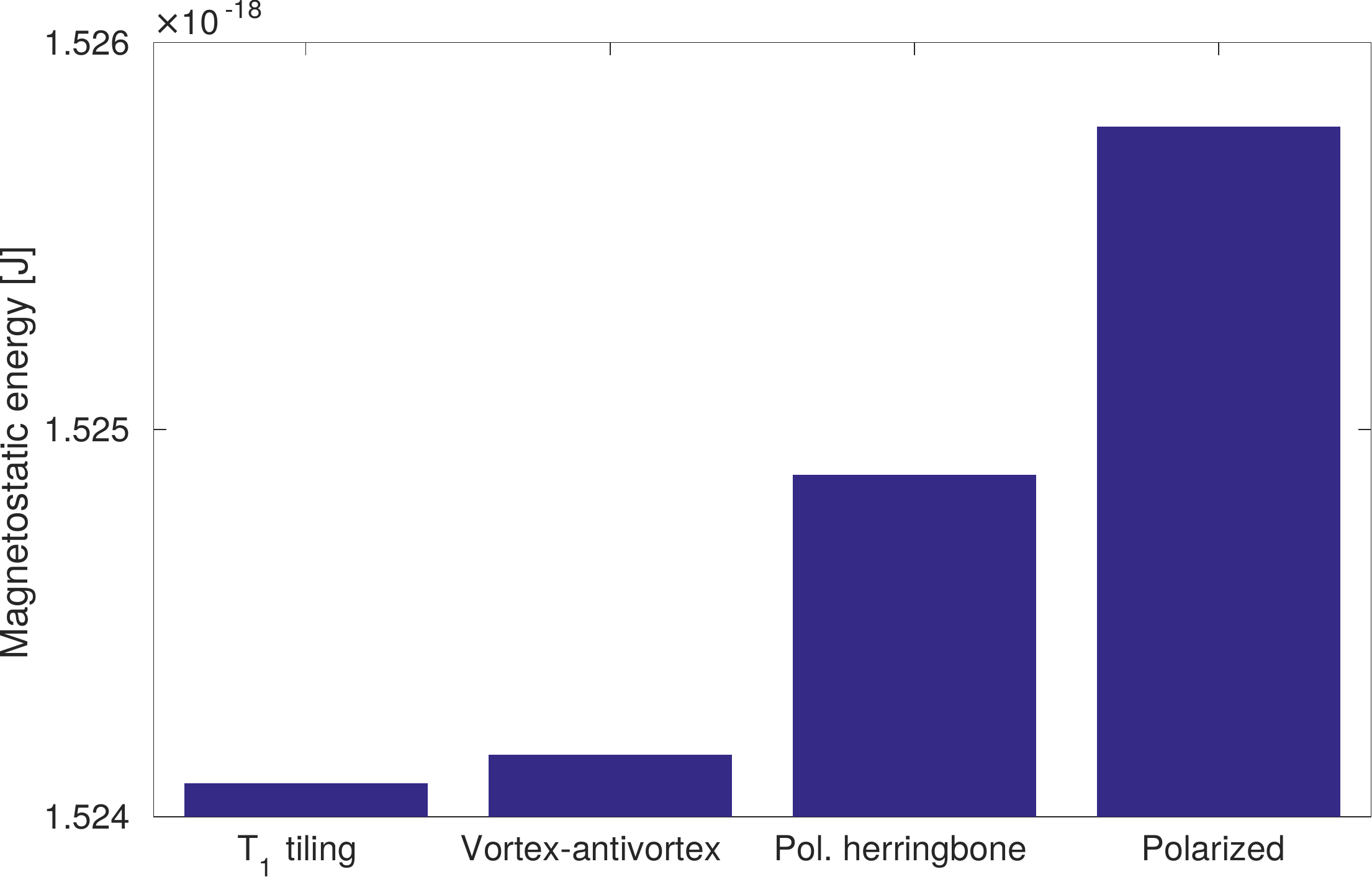}
\caption{\label{fig:a12} Energy comparison between the different all $\text{T}_\text{II}$ states, see Fig. 4, calculated using  M{\scriptsize U}M{\scriptsize AX}3 as defined in the paper.}
\end{figure}

\begin{figure}[h]
\includegraphics[width=0.5\columnwidth]{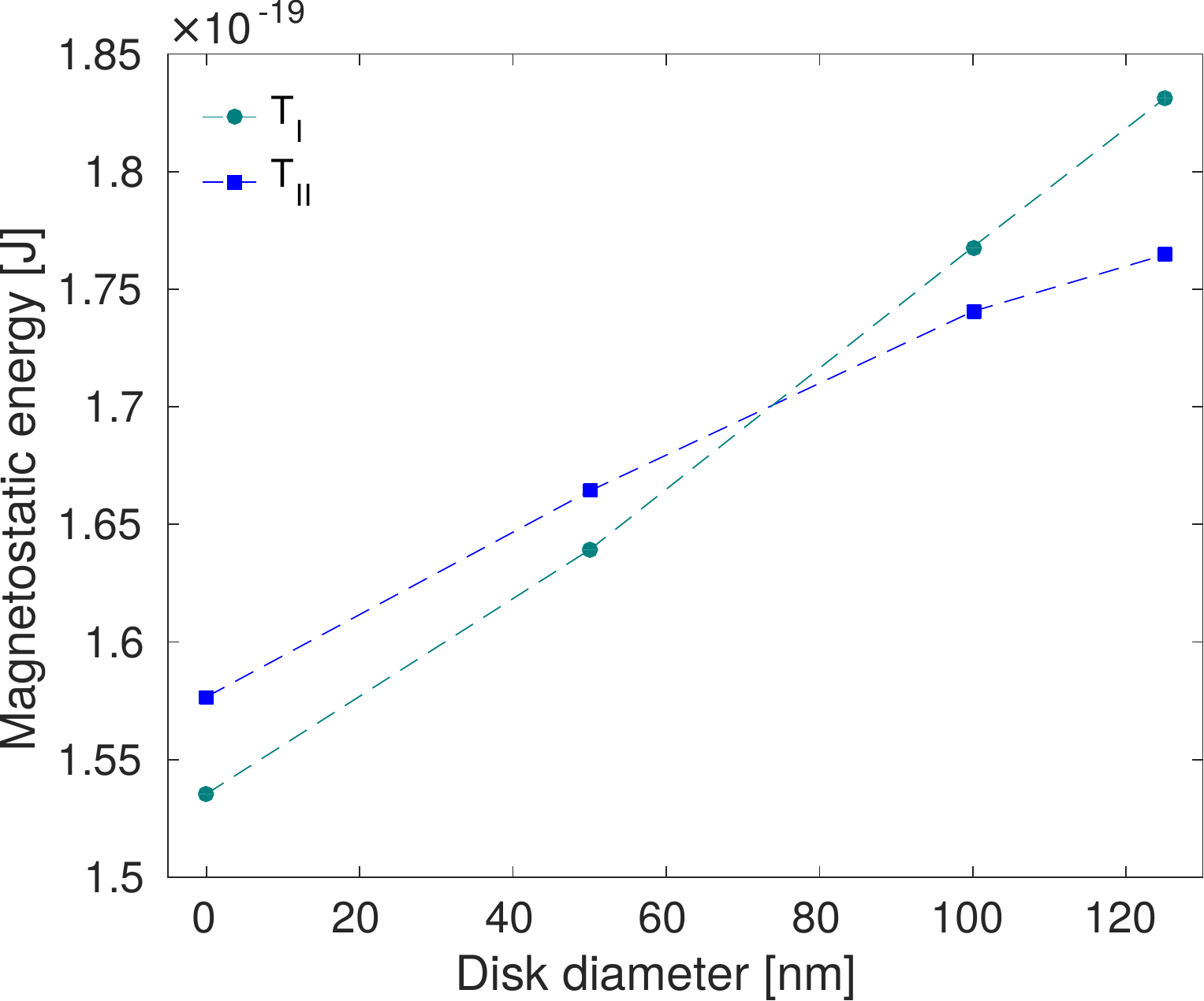}
\caption{\label{fig:a13} Energy comparison between $\text{T}_\text{I}$ and $\text{T}_\text{II}$ vertices for $\alpha$=660 nm with increasing disc diameter calculated using  M{\scriptsize U}M{\scriptsize AX}3. The crossover in energy happens at smaller disc diameters then what is suggested by the population inversion see Supplementary Figure \ref{fig:a1}. It is not feasible for a micromagnetic simulation to capture all the physics contained in a thermal system and it is therefore not surprising that the 0 K simulation do not quantitatively match the experiment to full extent.}
\end{figure}

\begin{figure}[h]
\includegraphics[width=0.5\columnwidth]{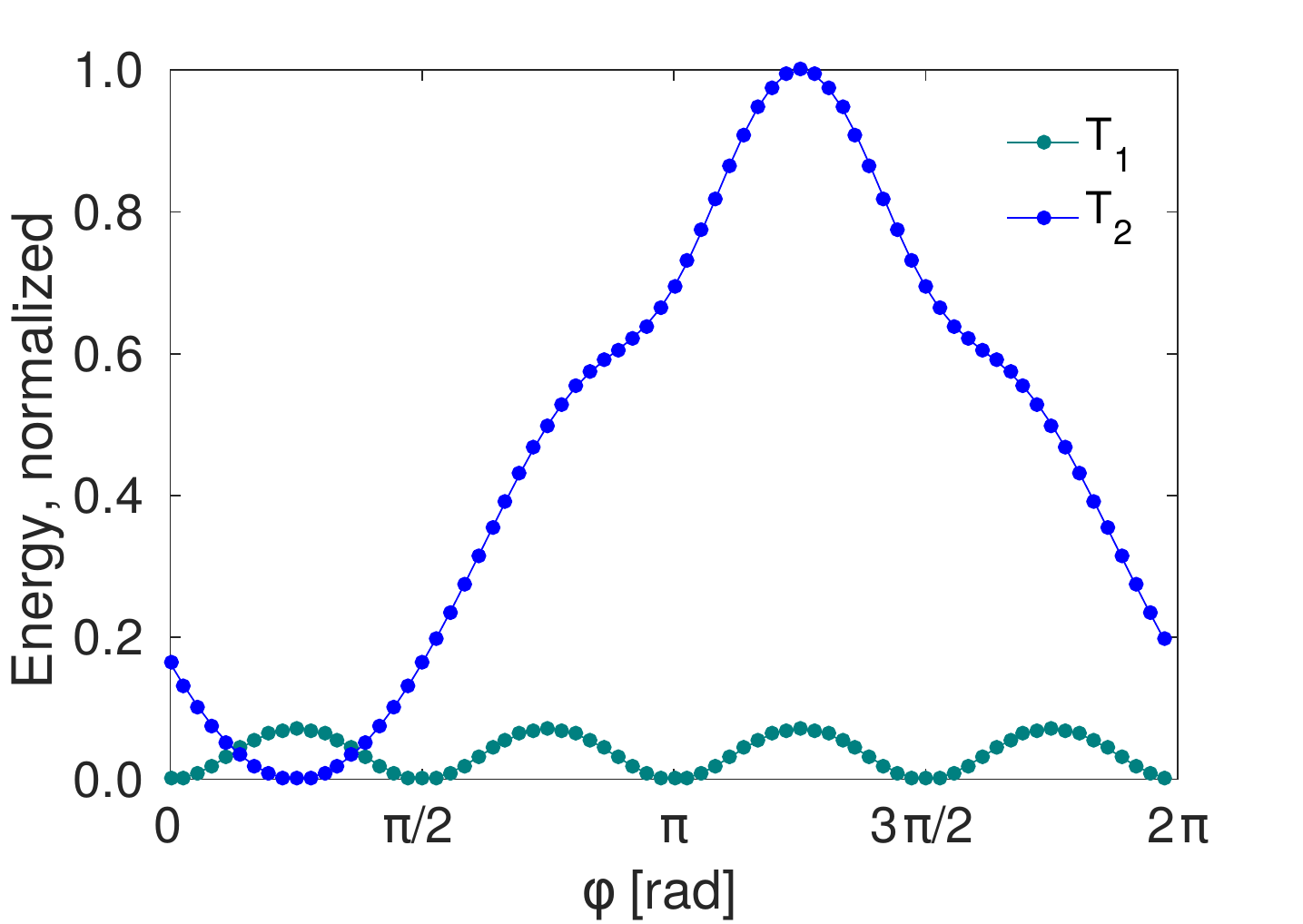}
\caption{\label{fig:a14} Energy for $\text{T}_\text{I}$ and $\text{T}_\text{II}$ vertices in a purely collinear XY-spin approximation for the disc's magnetization while rotating the magnetization of the disc. The energies are normalized to the maximum energy of the T$_{\text{II}}$ vertex. The energy minima for the disc in $\text{T}_\text{I}$ vertices is four fold (the same is true for $\text{T}_\text{IV}$) vertices). For $\text{T}_\text{II}$ and $\text{T}_\text{III}$ vertices (not shown) there is one energy minima along the combined flux of the islands. Compare to Fig. 1 in the paper. $\varphi$=0 rad lies along the long axis of one island.}
\end{figure}

\begin{figure}[h]
\includegraphics[width=0.8\columnwidth]{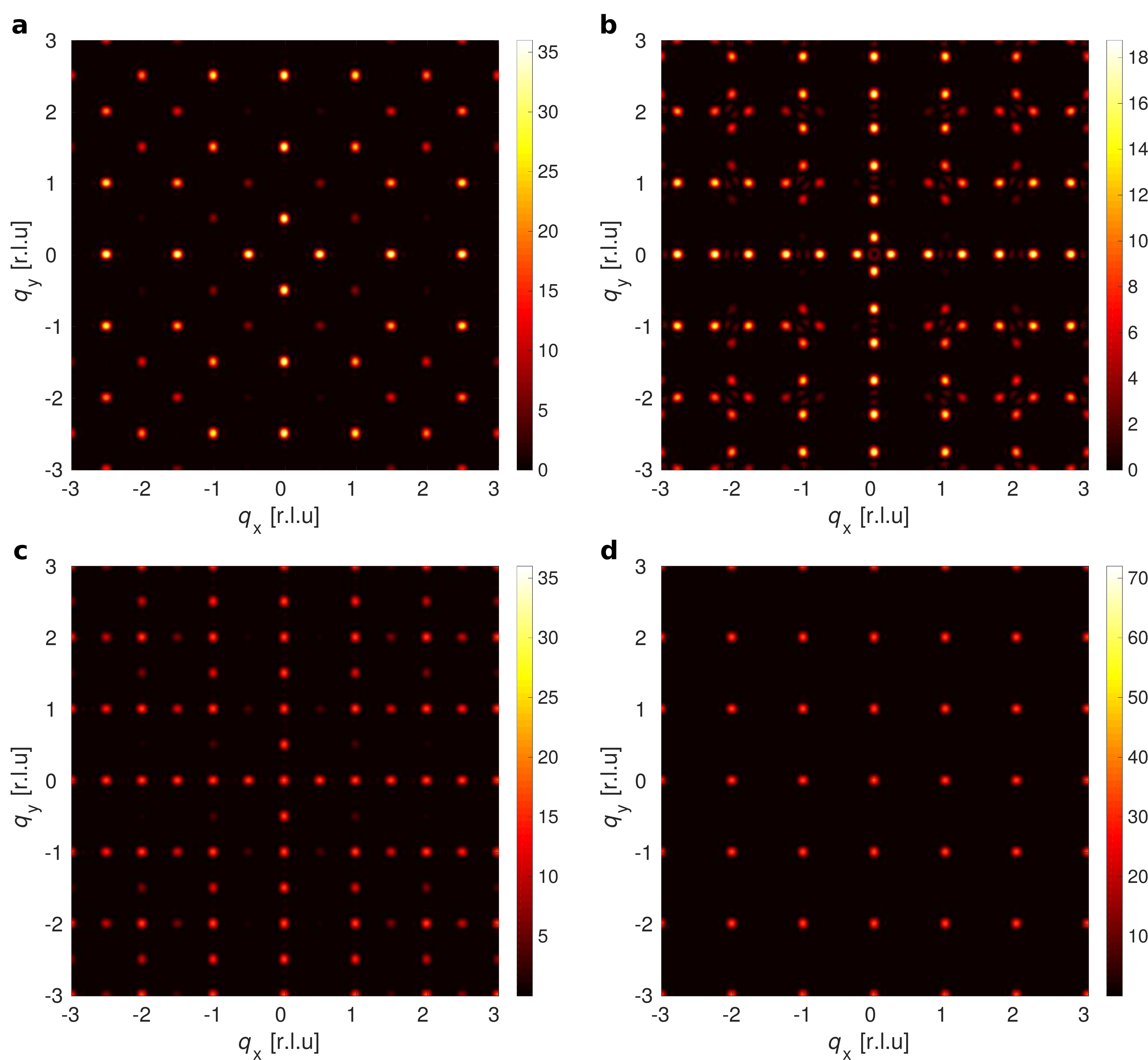}
\caption{\label{fig:a15} Computed magnetic spin structure factor for the different all $\text{T}_\text{II}$ states, see Fig. 4 in the paper. \textbf{a} $\text{T}_\text{I}$-tiling, \textbf{b} vortex-antivortex, \textbf{c} herringbone structure, and \textbf{d} is the polarized state. The points at $[q_x,q_y]$=[n$\pm$0.25,0] or $[q_x,q_y]$=[0,n$\pm$0.25], where n is an integer, are unique to the vortex-antivortex state, \textbf{b}. Peaks at integer values are shared between \textbf{c} and \textbf{d}. Points at half integer values are shared between \textbf{a} and \textbf{c}.} 
\end{figure}

\clearpage
\bibliographystyle{natphys}
\bibliography{mylib}